\documentclass[12pt]{article}
\usepackage{amsmath}
\newcommand{\bee}{\begin{eqnarray*}}
\newcommand{\ene}{\end{eqnarray*}}
\newcommand{\beeq}{\begin{equation}}
\newcommand{\eneq}{\end{equation}}
\newtheorem{lem}{Lemma}[section]
\newcommand{\bel}{\begin{lem}}
\newcommand{\enl}{\end{lem}}
\newtheorem{defi}{Definition}[section]
\newcommand{\bef}{\begin{defi}}
\newcommand{\enf}{\end{defi}}
\newtheorem{exap}{Example}[section]
\newcommand{\beex}{\begin{exap}}
\newcommand{\enex}{\end{exap}}
\newtheorem{theo}{Theorem}[section]
\newcommand{\beth}{\begin{theo}}
\newcommand{\enth}{\end{theo}}
\newtheorem{prop}{Proposition}[section]
\newcommand{\bep}{\begin{prop}}
\newcommand{\enp}{\end{prop}}
\newtheorem{cor}{Corollary}[section]
\newcommand{\bec}{\begin{cor}}
\newcommand{\enc}{\end{cor}}
\newtheorem{rem}{Remark}[section]
\newcommand{\ber}{\begin{rem}}
\newcommand{\enr}{\end{rem}}

\usepackage{amsmath}
\usepackage{graphicx}
\usepackage{latexsym}
\usepackage{xcolor}
\usepackage{latexsym}
\usepackage{graphicx}
\usepackage{rotate}
\usepackage{lscape}
\usepackage{float}

\begin{document}

\title{
FIDUCIAL MATCHING
\\  FOR\\  THE 
APPROXIMATE POSTERIOR: F-ABC
 }

\author {Yannis G. Yatracos, Yau Mathematical Sciences Center, Tsinghua University
}

\maketitle

 \vspace{4 in}

{\it Some key words:} \quad  Approximate Bayesian Computation; Bayesian consistency;
Fiducial (F)-ABC; F-ABC for all;  Matching Support Probability for the Tolerance; Samplers

\pagebreak

\pagebreak

\begin{center}
\vspace{0.1in} {{\large Summary}}
 \end{center}
\parbox{5.2in}

{\quad Approximate Bayesian Computation (ABC) provides    posterior models for  a stochastic parameter $\Theta$
when  the observed, 
size $n$ 
sample ${\bf x} (\in R^{nxd}),$ has intractable likelihood; $d \ge 1.$
For ${\bf x}$ from  c.d.f. $F_{\theta},$   with unknown $\theta \in {\bf \Theta},$  the ABC-steps are: a {\em single} sample, ${\bf x}^*,$ is
drawn  from $F_{\theta^*},$  with known $\theta^* \in {\bf \Theta};$ 
a {\em nearly}  sufficient  
summary,  
$S({\bf x}),$  is {\em determined}
for matching  ${\bf x}$ and ${\bf x}^*$ within  {\em some} $\epsilon$-tolerance for  
{\em a}  distance-measure $\rho$; if ${\bf x}$ and  ${\bf x}^*$ match,  $\theta^*$ is included
  in the approximate posterior  with weight, $K({\bf x}, {\bf x}^*;\epsilon); K$ is 
 {\em  arbitrary  kernel}.
We introduce 
Fiducial (F)-ABC,  with 
$M$ ${\bf x}^*$  drawn from $F_{\theta^*}.$ The  goal is  a  ``one-for-all F-models'' approach,
with $\theta^*$-weights not $K$-artifacts and 
 with universal sufficient $S$ the empirical measure, $\mu_{\bf X},$ when $d>1,$ 
 the empirical cumulative distribution, $\hat F_{\bf X},$ when  d=1, and, respectively   
for  $\rho,$ the Total-Variation 
and the Kolmogorov distance, $d_K.$
Light is thrown to
$\epsilon$'s nature
via $d_K,$  guidelines are given to determine its value and the  
``0-1'' restrictive influence on $\theta^*$  is reduced.
$\theta^*$-weight   
is the proportion of ${\bf x}^*$ 
matching ${\bf x}$  which, for many  models, 
  increases to 1 as $\theta^*$ converges  to $\theta,$
unlike $K({\bf x}, {\bf x}^*;\epsilon).$
The number of simulations
for  implementation is moderate. 
Under few, mild assumptions,  F-ABC posterior converges  to its target and
rates of concentration to $T(F_{\theta})$  are obtained; $T$ functional.
When $M=1,$ F-ABC is reduced to ABC. 
When $M>1,$  F-ABC posterior  includes either selected $\theta^*$  or 
{\em all}  $\theta^*$ used,  reducing $\epsilon$'s influence. 
In simulations, nonparametric F-ABC posterior improves  the concentration of {\em parametric}  ABC posterior  at $\theta$
and ``F-ABC posterior for all $\theta^*$'' is satisfactory.\\



\pagebreak

 \section{\bf Introduction}

\quad  In Bayesian inference, central theme is
the posterior model, $\pi(\theta^*|{\bf x}),$  of  stochastic parameter $\Theta$
given the observed data, sample ${\bf x}; \theta^* \in {\bf \Theta}.$ 
Approximate Bayesian Computation (ABC) method 
 provides {\em a}   posterior model  
 when the data's likelihood is intractable.
Rubin (1984) described  the first  ABC method for 
${\bf x}$ from the model  
$f({\bf y}|\theta)$ with cumulative distribution function ({\em c.d.f.}) $F_{\theta}({\bf y}),$
using simulated ${\bf x}^*$-samples
for  several $\theta^*$-values 
 having each 
$\Theta$-prior $\pi(\theta); \theta, \theta^*  \in {\bf \Theta}, {\bf x}, {\bf x}^*  \in R^{nxd},$ generic sample value  ${\bf y} \in R^{nxd}, d \ge 1, n$ is the sample size.  The 
$\theta^*$ for which 
${\bf x}^*$ ``matches'' (or ``looks similar to'')
${\bf x}$ 
 are 
$\Theta$'s approximate
posterior. 

Since then, several  research results have been obtained in  ABC,
creating  the new statistical culture of  Bayesian-Frequentists.
Robert (2017) provides a  survey on recent ABC results, including 
  three 
approximations/concerns:\\
{\em i)} ABC degrades the data precision down to a tolerance level $\epsilon,$ replacing
the event 
${\bf X}={\bf x}$ with the event
\begin{equation}
\label{eq:topol} 
\rho({\bf X}^*, {\bf x} ) \le \epsilon;
\end{equation}
$ \rho$ is a distance-measure.\\
{\em ii)} ABC substitutes for the likelihood a non-parametric approximation.\\
{\em iii)} ABC summarizes
${\bf x}$
 by {\em an almost always insufficient statistic}, $S({\bf x}),$ 
using instead of (\ref{eq:topol}),
\begin{equation}
\label{eq:topolS} 
\rho(S({\bf X}^*), S({\bf x}) ) \le \epsilon.
\end{equation}

The basic ABC-rejection algorithm selects $\theta^*$ when  either (\ref{eq:topol}) or  (\ref{eq:topolS}) holds (Tavar\'{e} {\em et al.}
1997, Pritchard {\em et al.}, 1999). Recently, ${\bf x}^*$ are drawn from a Sampler.

There are additional concerns on ABC.
{\em a)} The dimension of
$S$ with Big Data
when the statistical nature of $\theta$ is unknown.
{\em b)} The $\epsilon$-value used and  the ``0-1''  restrictive influence on $\theta^*,$ 
$\epsilon$'s  missing  sampling interpretation and  components,
$\epsilon$'s dependence on
$n$ and  the  distance between $F_{\theta}$ and selected $F_{\theta^*}.$
{\em c)} The acceptable number of $\theta^*$ in the posterior.
{\em d)} For continuous ${\Theta},$  
 the  arbitrary weights ``0'' or ``1'' or $K(\frac{{\bf x}-{\bf x}^*}{\epsilon}),$  given 
to $\theta^*$  at {\em any} distance from $\theta,$ using
 the  ``one and only'' ${\bf x}^*$  from {\em c.d.f.} $F_{\theta^*};$ 
 $K$ is an arbitrary  kernel. {\em e)} The $\theta^*$-weights in the approximate posterior create a $K$-dependent artifact; $K$ is usually a normal kernel.
{\em f)} It is not clear whether {\em non-selected} $\theta^* (\in R)$ 
 is included in the approximate  posterior when $\theta_1^*, \theta_2^*$ are selected and $\theta_1^*<\theta^*<\theta_2^*.$
{\em g)} For discrete $\Theta$ and  with $\theta^*$ drawn twice,  it is not clear whether
$\theta^*$ is  selected if only one of the simulated ${\bf x}_1^*, {\bf x}_2^*$ matches ${\bf x}.$  
{\em h)} Pure Bayesians and frequentists  may question the $\epsilon$-exclusion of non-selected $\theta^*$ from the approximate posterior.

Bernton {\em et al.} (2019)\footnote{The details are given for Editors, AEs, referees and readers to avoid confusion, but could be reduced.} propose to solve the choice-problem for 
$S$ and $\rho$   using, respectively, the ``empirical distribution'' with ``abuse of language'' (section 1.1, 1st paragraph)
and (basically)  Wasserstein distance, $d_{W}.$ 
The latter is computed for the observed ${\bf x}$ and the  ``synthetic'' data, 
${\bf x}^*,$ and is baptized ``distance between empirical distributions'' when introducing (5),
but no ``empirical distributions'' appear in $d_W$  even though used in statements.
In the abstract, it is stated that the approach avoids ``the use of summaries and the ensuing loss of information by instead using the Wasserstein distance
{\em between the empirical distributions} of the observed and synthetic data.'' but in section 1.3, first paragraph, it is instead stated ``{\em hoping  to avoid} the loss of information incurred by the use of summary statistics''. The authors associate ``no information loss'' with the case $d_W({\bf x}, {\bf x}^*)=0$ without examining whether there is information loss when $d_W({\bf x}, {\bf x}^*)$ is smaller than  positive $\epsilon.$
For $d_W$ that ``metrizes empirical distributions'' 
consider, for example, the extreme case of only one observation, $n=1,$
the observed data ${\bf x}=(x_1,\ldots,x_d),$  the synthetic  
${\bf x}^*=(x_1- \epsilon, \ldots ,x_d- \epsilon) $   and $\delta_{\bf x}, \delta_{{\bf x}^*} $ the corresponding Dirac functions,with  $\epsilon>0$ much smaller in magnitude than all ${\bf x}$ coordinates.  Using $d_W$ that is sum of absolute differences of the coordinates of ${\bf x}$ and ${\bf x}^*,$ the
 distance 
$d_W({\bf x}, {\bf x}^*)=d \cdot\epsilon$ but the Kolmogorov distance $d_K$ and the  Total Variation distance, TV, between 
$\delta_{\bf x}$ and $\delta_{{\bf x}^*} $ take
their  maximum value  1, and this holds for  any $\epsilon$ decreasing to zero. Similar results can be obtained for  fixed size $n$ samples 
${\bf x}$ and ${\bf x}^*$ in $R^d,$ with  same form.
Then, $d_K(\delta_{\bf x}, \delta_{{\bf x}^*})/d_W({\bf x}, {\bf x}^*)$ and 
$TV(\delta_{\bf x}, \delta_{{\bf x}^*})/d_W({\bf x}, {\bf x}^*)$ both diverge to infinity as $\epsilon$ converges to zero and $d_W$ is not equivalent to $d_K$ and $TV,$ leading to different neighborhoods and convergence. By definition, the $d_K$ and $TV$
values for empirical cumulative distribution 
functions  and empirical measures, respectively, are bounded by 1 but this does not always  hold for $d_W({\bf x}, {\bf x}^*).$ 
Summarizing, loss of information remains when using $d_W$ and neither empirical distributions, nor empirical measures are used in calculations. The authors 
avoid the word ``sufficiency'' in the paper, which is also indicative of potential  information loss.

Loss of information due to summary statistics or a method, e.g., Bernton {\em et al.} (2019),  can be avoided using the empirical 
cumulative distribution function, $\hat F_{\bf X},$ when $d=1$ and the empirical measure, $\mu_{\bf X}(A),$ when $d>1; A \in {\cal B}_d,$ the Borel sets in $R^d, d>1.$ To match ${\bf x}$ with ${\bf x}^*, d_K(\hat F_{\bf x}, \hat F_{{\bf x}^*})$ is used when $d=1.$ 
 When $d>1,\mu_{\bf X}$ and $\mu_{{\bf X}^*}$  need to be compared on  every Borel set to measure the information difference, and the supremum of the absolute differences over all Borel sets provides the maximum information loss.
This is Total Variation (TV) distance
to be used as  $\rho$-distance herein. TV has the advantage of  matching and separating well probabilities $P$ and $Q$ in $R^d,$ which  are equal when $P(A)=Q(A)$ for every $A \in {\cal B}_d,$ and is useful for $\epsilon$-matching $\mu_{\bf x}$ with $\mu_{{\bf x}^*}.$

Concerns {\em a)}-{\em h)} inspired also
the search for an alternative to ABC. The $S$ and $\rho$-choices are:   
$\mu_{\bf X}$   with the TV-distance,  when $d>1;$ 
 $\hat F_{\bf X}$  with the Kolmogorov distance, $d_K,$  when $d=1$ (see section 5).
Motivated by the Conditional Calibration framework 
(Rubin, 2019) and 
a phenomenon observed  in several models,
the Fiducial (F)-ABC  matching is introduced,
supported by 
$M$  ${\bf x}^*$   drawn from  $F_{\theta^*}, 50\le M \le 200.$ 
 $p_{match}(\theta^*)$ is the   ${\bf x}^*$-proportion 
within the $\epsilon$-tolerance, used as $\theta^*$-weight in
the F-ABC posterior.
 $p_{match}(\theta^*)$  estimates   the {\em ${\bf x}^*$-matching support probability} $\alpha$ of  event
(\ref{eq:topolS})
that provides $\epsilon$'s sampling interpretation and  value; $0\le \alpha \le 1.$
 In practice,  $\epsilon$ is determined  via $\alpha$ and  the Sampler (section \ref{s:epsilonand1-alpha}).
For  several $F_{\theta^*}$-models,  $p_{match}(\theta^*)$ 
converges to 1 as $\theta^*$ converges to $\theta,$ unlike $K(\frac{{\bf x}-{\bf x}^*}{\epsilon}).$

In ``F-ABC for all''  each drawn $\theta^*$ is included in  the posterior with weight $p_{match}(\theta^*),$
reducing $\epsilon$'s influence and without using a kernel. When $M=1,$  F-ABC
is
ABC.
The use of $M$ ``pseudo-samples'' is non-traditional (see, {\em e.g.}  Bornn {\em et al.}, 2017, and references therein), 
extracting with $p_{match}(\theta^*)$ useful $\theta$-related information for the posterior.
 The $\theta^*$-value  maximizing $p_{match}(\theta^*)$  is the Maximum Matching Support Probability Estimate (MMSPE, Yatracos, 2020).




Simulations indicate that 
{\em nonparametric} F-ABC competes well with  parametric, flat-kernel ABC and improves 
very frequently  the concentration of the approximate posterior. The  graphs of the F-ABC posterior 
for all $\theta^*$ drawn
should not pass unnoticed since the Bayesian posterior, $\pi (\theta^*|{\bf x}),$  is inclusive of {\em all} $\theta^*$ with different weights.
 
For the ${\bf X}^*$-matching 
support probability $\alpha$ with $\rho=d_K,$ an upper bound $\epsilon_{n,B}$ on $\epsilon_n$ is determined;  $0<\alpha <1.$
$\epsilon_{n, B}$  has two additive components:  
{\em I)}  the observed or acceptable discrepancy  between $F_{\theta}$ and  the  $F_{\theta^*}$-models, and
{\em II)}  a component  determined by a confidence 
related to $\alpha$ (section \ref{sec:study epsilon}). 
Under exchangeability on $F_{\theta}({\bf y}),$  the ABC and F-ABC posteriors with $d_K$-matching   converge  to 
$\pi(\theta|{\bf x})$ when
$\epsilon$
converges to zero; $n$ is  fixed.  For a continuous linear functional $T$ on the space of {\em c.d.fs}, Bayesian consistency is established  and the rate of concentration of $T(F_{\theta^*})$ around $T(F_{\theta})$ depends on $\epsilon_n,$ the rate of concentration in probability  of $\hat F_{\bf X}$ around $F_\theta,$
and $T$'s modulus of continuity (section \ref{sec:asy}).



 Lintusaari {\em et al} (2017) and Fearnhead (2018) provide
accessible
 introductions to ABC presenting, respectively,  recent developments and 
results on asymptotics. 
Tanaka {\em et al.} (2006, p. 1517 and Figure 4) indicate $\epsilon$'s choice   is crucial for the sampler acceptance rates and the posterior densities.
 Fearnhead and Prangle (2012) show how to construct appropriate summary 
$S$ for ABC
to be used in  (\ref{eq:topolS}) and enable inference about $\theta.$
Frazier {\em et al.}  (2015) derive conditions under which $S$ yields consistent Bayesian inference.
Biau {\em et al.} (2015),  analyze  ABC as a  $k$-nearest neighbor method. 
 Frazier {\em et al.} (2018) provide 
for the posterior: its  concentration rate  on sets containing $\theta,$
 its limiting shape and the asymptotic distribution of its mean.
Nott {\em et. al.} (2018) approximate 
Bayesian predictive $p$-values with Regression ABC.
 Vihola and Franks (2020) 
suggest  a balanced $\epsilon$  from a {\em range}  of tolerances via Bayesian MCMC.

When  ${\bf x}$ is not
 obtained from 
models 
$\{f({\bf y},\theta) \pi(\theta), \theta \in {\bf \Theta}\}$ but ${\bf X}^*$ is, 
Miller and Dunson (2019) propose a robust ABC approach conditioning on  $\rho$-neighborhoods of 
empirical {\em c.d.fs} $\hat F_{\bf x}$ and $\hat F_{{\bf X}^*},$
 suggesting among  $\rho$-distances  
$d_K$ (for real valued observations only), but  use  Kullback-Leibler divergence
for their  ABC coarsened({\em c})-posterior. 


\section{Fiducial ABC}

\quad Let $\pi(\theta)$ be the prior for $\Theta$  with respect to measure $\nu$ on 
${\bf \Theta}$ with $\sigma$-field ${\cal C}_{\bf \Theta}, \theta \in {\bf \Theta}. \  {\bf y}$ is generic sample value. ${\bf X}$ is  a sample  of size $n$ obtained 
 from the unknown $\theta$-model with cumulative distribution function $F_{\theta}({\bf y})$ and density $f_{\theta}({\bf y})$ (or
$f({\bf y}|\theta)$)  with respect to  measure $\mu$ on ${\cal Y}$  with $\sigma$-field ${\cal C}_{\cal Y}.$
$ {\cal Y}$ is usually subset  of  $R^d$ with the Borel $\sigma$-field, ${\cal B}_d,  d \ge 1.$
$ \pi(\theta|{\bf y})$ is the posterior of $\Theta. \   {\bf X}^*$ is a sample of size $n$ obtained from the sampler 
 with model  $F_{\theta^*}.$ 
$S({\bf X})$ is  a summary for ${\bf X},$ 
$\rho$ measures the distance between $S({\bf x})$ and $S({\bf X}^*).$ As statistic $S({\bf X})$ can be thought of as estimate
of $T(F_{\theta}), \ T$ generic functional of $F_{\theta}.$ 
For $A \in {\cal C}_{\bf \Theta} (\mbox{ or }{\cal C}_{\cal Y}), I_A({\bf u})=1$ if ${\bf u} \in A$ and zero otherwise.
${\bf \Theta}$ is metrized with $d_{\bf \Theta}$ and generic $\tilde d$ and $d_K$ are distances for c.d.fs.
$\theta$-identifiability is assumed, {\em i.e.}, $F_{\theta_1}=F_{\theta_2}$  implies $\theta_1=\theta_2.$

 
\bef
\label{d:MSP}
For tolerance $\epsilon,{\bf X}$ and $S,$  the ${\bf X}^*$-matching support probability $\alpha$ for $\theta^*$ is
\beeq
\label{eq:tolfabcforone}
P[\rho(S({\bf X}^*), S({\bf X} )) \le \epsilon]= \alpha, \ 0 \le \alpha \le 1.
\eneq
Given 
$\epsilon>0$ and ${\bf \Theta}^*=\{\theta_1^*, \ldots, \theta_N^*\},$ the matching support probability for ${\bf \Theta}^*$
is 
\beeq
\label{def:MSPgroup}
\inf \{\alpha_i; i=1,\ldots, N\};
\eneq
\enf
$\alpha_i$ is obtained from (\ref{eq:tolfabcforone}) for $\theta^*=\theta_i^*, i=1,\ldots,N.$

The probability in 
(\ref{eq:tolfabcforone})
is not under one probability model as in confidence band calculations since  ${\bf X}$ and ${\bf X}^*$ follow 
$F_{\theta}$ and $F_{\theta^*},$ respectively.
When ${\bf X}={\bf x},$  
$\epsilon$ is the 
$\alpha$-quantile of $\rho(S({\bf X}^*), S({\bf x} ))$ under $F_{\theta^*}$  and seeing density as ``small probability", 
\beeq
\label{eq:whypmatch}
\pi(\theta^*|{\bf x})\propto f({\bf x}|\theta^*) \propto P_{\theta^*}[\rho({\bf X}^*,{\bf x}) \le \epsilon],
\eneq
for small $\epsilon,$ used in (\ref{eq:moregeneralthantopol}) with $S({\bf x)}$ instead of ${\bf x}.$
The $\alpha$-value 
is omitted from the notation F-ABC since it will be determined in the Algorithm, along with $\epsilon.$ 



F-ABC  {\em Algorithm}
\mbox{ }\\
{\bf 1)} {\em Determination of $\epsilon_n, \alpha_n:$ \footnote{We consider it part of the algorithm due to repeated samples from $F_{\theta^*}.$ If referees prefer it separated, the change will be made.}}  
Sample  several $\theta^*$-values either from $\pi(\theta)$ or from
a discretization of $\bf \Theta$ if it is known. Use one of them as base-value,  $\theta^*_b,$ and obtain ${\bf x}$ generated by 
$\theta^*_b.$ {\em Select}, e.g., 5-10 $\theta^*$ at increasing {\em standardized} distance from $\theta^*_b$ taking into consideration its nature and obtain $M$ ${\bf X}^*$-samples from
each one of them and $\theta^*_b.$ Calculate $\rho(S({\bf X}^*_i), S({\bf x})), i=1,\ldots, M,$
and their empirical quantiles for each one of the selected $\theta^*$ and $\theta^*_b.$ 
Create a table similar to  Table 1 in subsection \ref{s:epsilonand1-alpha}.
 After consultation of the quantiles decide on the $\epsilon_n$ 
to be used, determined from $\theta^*_{\epsilon_n}$ with corresponding quantile $\alpha_n.$\\
{\bf 2)} Sample {\em i.i.d.} $\theta_1^*, \ldots, \theta_{N^*}^*$ from ${\bf \Theta}$ according to $\pi(\theta).$\\
 {\bf 3)} Repeat for $i=1,\ldots, N^*;$ F-ABC is potentially  used for all $\theta_1^*, \ldots, \theta_{N^*}^*.$ \\
{\em a)} Sample ${\bf X}_1^*, \ldots, {\bf X}_M^*$ from $f({\bf y}|\theta^*_i).$\\
{\em b)} Compute the observed matching support proportion, $p_{match}(\theta_i^*),$  for  the ${\bf x}_1^*, \ldots, {\bf x}_M^*:$ 
\begin{equation}
\label{eq:moregeneralthantopol} 
p_{match}(\theta_i^*)=\frac{Card(\{{\bf x}_i^*:\rho(S({\bf x}_i^*), S({\bf x} )) \le \epsilon_n, \ i=1,\ldots, M\})}{M}. 
\end{equation}
 {\em c)}  {\em  $\theta^*$-selection criterion: {the \em{F-ABC}} filter.}\footnote{Not used in F-ABC for all $\theta^*.$
It is intended for users desiring to restrict further the approximate posterior.}  Include $\theta^*_i$ in the domain of $\pi(\theta|{\bf x})$ when
\begin{equation}
\label{eq:thresholdcriterion} 
p_{match}(\theta_i^*) \ge \alpha_n.
\end{equation}
{\bf 4)} The selected $\theta^*$ in {\bf 3)}  after the end of the algorithm are 
\beeq
\label{eq:selectedn}
{\bf \Theta}_n^*=\{\theta_{sel, i}^*; i=1,\ldots, N\}, \ N\le N^*.
\eneq
Use $\{(\theta_{sel, i}^*, p(\theta_{sel, i}^*)); i=1,\ldots, N\}$ to construct  the F-ABC posterior.



\bef
\label{def:observedmatchingsupportprob}
For ${\bf \Theta}_n^*$ in (\ref{eq:selectedn}) the observed matching support probability is $\min \{ p(\theta_{sel, i}^*); i=1,\ldots, N\}.$
\enf

\ber
\label{r:ABCcompFABC} Comparing ABC with F-ABC:  When $M=1$ in  {\bf 3)}a) and $\alpha_n=1$ in (\ref{eq:thresholdcriterion}), $\rho${\em-F-ABC} is
$\rho${\em-ABC}.
To compare $\rho_1${\em-ABC} with $\rho_2${\em-F-ABC}, start with $\rho_2${\em-ABC}, use $M$ additional 
${\bf x}^*$-samples for the selected $\theta^*$ to obtain $p_{match}(\theta^*)$ for all $(M+1)$ ${\bf x}^*$-drawn, 
and proceed with {\bf 4)} to construct the $\rho_2${\em-F-ABC} posterior. When $\alpha_n=0$   in (\ref{eq:thresholdcriterion}),
all $\theta^*$ are selected for the posterior with their corresponding weight, $p_{match}(\theta^*).$
 \enr  


Let  
\beeq
\label{eq:B}
B_{\epsilon_n}= \{  {\bf x}^*: \rho(S({\bf x}^*), S({\bf x} )) \le \epsilon_n\},
\eneq
without specifying the values of $\alpha$ and $M$ which will be determined by the context. Similarly,
  the F-ABC posterior of theta  is 
\beeq
\label{def: f-abcposterior}
\pi_{f\mbox{-}abc}(\theta|B_{\epsilon_n})
=\frac{\pi(\theta) \cdot
\int_{\cal Y}  I_{B_{\epsilon_n}}({\bf y})  f({\bf y}|\theta) \mu(d{\bf y})}
{\int_{\bf \Theta}  \pi(s)  \int_{{\cal Y}}I_{B_{\epsilon_n}}({\bf y}) f({\bf y}| s) \mu(d{\bf y})  \ \nu(ds)    },
=\frac{\pi(\theta) \cdot P_{\theta}^{(n)}(B_{\epsilon_n})}{\int_{\bf \Theta}\pi(s) \cdot P_{s}^{(n)}(B_{\epsilon_n}) \nu(ds)}.
\eneq
and for $H \in {\cal C}_{\bf \Theta},$ its F-ABC probability is
\beeq
\label{eq:fabcposteriorProb}
\Pi_{f\mbox{-}abc}(H|B_{\epsilon_n})=\int_H  \pi_{f\mbox{-}abc}(\theta|B_{\epsilon_n}) \nu(d\theta)
=\frac{\int_{\bf \Theta}\pi(\theta) \cdot P_{\theta}^{(n)}(H\cap B_{\epsilon_n })\nu(d\theta)} {\int_{\bf \Theta}\pi(s) \cdot P_{s}^{(n)}(B_{\epsilon_n}) \nu(ds)}.
\eneq
For ABC, $\pi_{abc}$ and $\Pi_{abc}$ are used instead.




\bef
\label{def:d_K}
 For any two distribution functions $F, G$ in $R^d, d\ge 1,$  their Kolmogorov distance
\begin{equation}
\label{eq:Kolmogorov}
d_K(F,G)=\sup \{|F(y)-G(y)|;  y \in R^d\}.
\end{equation}
\enf

\bef
\label{eq:def:empiricalcdf}
For any $n$-size sample
${\bf Y}=( Y_1,\ldots, Y_n) $
of  random vectors in 
$R^d, n\hat F_{\bf Y}(y)$
denotes the number of 
${Y}_i$'s with
all their components  smaller or equal to the corresponding components of 
$y. \  \hat F_{\bf Y}$ is the empirical c.d.f. of ${\bf Y}.$
\enf

In section \ref{s:implementation}, for observations in $R$ use in {\bf 1)} of the F-ABC Algorithm and 
 in (\ref{eq:moregeneralthantopol}): $S({\bf x})=\hat F_{\bf x}, \rho=d_K.$  
For observations in $R^d, d>1,  \hat F$ and $d_K$ will be used over 1-dimensional projections of the samples.
 


Implementation follows,  before the theoretical results for easier reading; could follow the theoretical results,  if required. 

\section{Implementation and Comparisons: ABC and F-ABC}
\label{s:implementation}

\quad The simulation results have no goal to compare for specific data sets F-ABC posteriors with W-ABC or ABC posteriors simply because the comparison does not make sense: F-ABC does not use an arbitrary chosen  Kernel, $K({\bf x}, {\bf x}^*;\epsilon),$ and has theoretical advantages with respect to ABC and W-ABC.
The simulations compare ABC with (${\hat F}_{\bf X}, d_K)$ and F-ABC with parametric ABC  to check the concentration of the posteriors and present posteriors created  without the use of Kernel, in particular histograms of the matching support probabilities, before using
the by default $R$-kernel for smoothing.
In Figures 1-3,  separate graphs are presented, mainly  for easier observation and  for not mixing domains and ranges of densities having an effect in plots.

 \subsection{
$\epsilon_n$ and matching support probability $\alpha$   in practice}
\label{s:epsilonand1-alpha}


\quad The goal is to 
implement the selection of $\epsilon_n$ and  $\alpha_n$  in ${\bf 1)}$ of the F-ABC Algorithm. When $\rho=d_K,$ upper bound
$\epsilon_{n,B}$ for $\epsilon_n$ is provided
in section \ref{sec:study epsilon}, but fine tuning is needed for  $\epsilon_{n,B}$  to be used  even for real observations. 
Bayesian-Frequentists 
and computer scientists  use efficiently  a  powerful tool: the sampler ${\cal M}$ for obtaining ${\bf X}^*$ from $F_{\theta^*}.$ 
 As illustration, Table 1  is provided for a sample of $n=100$   normal random variables with mean $\theta$ and
variance 1. With the notation in {\bf 1)} of F-ABC algorithm, $\theta^*_b=\theta=0$ and  ${\bf x}$ is obtained.
$M=500$  samples\footnote{$M=500>200$ to increase table's accuracy, with  execution time less than 15 seconds.}  are obtained for each 
$ \theta^*=0, (.5), 4$  and  $d_K$-distances are calculated; .5 corresponds to .5 standard deviation of the model.
 If $\epsilon=.63$  is used, it is expected that $\theta^*$ in the range $(-1.5, 1.5)$ are  selected and the 
observed matching support probability (Definition \ref {def:observedmatchingsupportprob})  will be (at least) .95. 
The dependence of $\epsilon$ and $\epsilon_{n,B}$ in the distance between $F_{\theta}$ and $F_{\theta^*}$ is confirmed.

\begin{table}[h!]
\begin{tabular}{|c|c|c|c|c|c|c|c|c|c|c|c|c|} \hline
\multicolumn{13}{|c|}
{\bf 
Empirical Quantiles of Kolmogorov distances between $\hat F_{\bf x}$  and  $\hat F_{{\bf x}^*}$} \\
\hline 
$\theta^*$ & MIN &  25th  & 50th & 60th & 65th & 70th & 75th & 80th & 85th & 90th & 95th & MAX \\ 
\hline
0  &  0.04   &  0.07 &   0.09 &   0.1 &    0.1 &   0.11 &  0.11 &  0.12 &  0.12  & 0.13 &   0.14 &  0.19\\ 
 \hline 
0.5  & 0.12  & 0.2 &  0.23 &  0.24  & 0.25 &  0.25 &  0.26 &  0.27 & 0.28 & 0.29 &  0.3  & 0.39\\    
\hline  
1  &   0.25 &  0.38 &   0.41  &  0.42 &   0.42 &   0.43 &   0.44  &  0.44  &  0.45 &   0.46 &  0.48 &  0.55\\  
\hline
1.5  & 0.47 &  0.55 &  0.57 &  0.58  & 0.59 &  0.59 &  0.6 &  0.61  & 0.61 &  0.62 & 0.63  & 0.69 \\
\hline
2  & 0.6 &  0.68 &  0.71  & 0.71 &  0.72  & 0.72 &  0.73 &  0.73 &  0.74 &  0.75 &  0.76 &  0.79\\ 
\hline  
2.5   &  0.72  &  0.8   & 0.82  &  0.83  &  0.83  &  0.83  &  0.84  &  0.84  &  0.85   & 0.86  & 0.87  & 0.91\\ 
\hline
3 &  0.82  & 0.89 &  0.9 &  0.91 &  0.91 &  0.91 &  0.92  & 0.92 &  0.92 &  0.93 &  0.93 &  0.95\\ 
\hline
3.5  &    0.89   &0.94   & 0.95  &  0.96  & 0.96  &  0.96  &  0.96  &  0.96  &  0.97  & 0.97  &  0.97  & 0.99\\
\hline
 4   &   0.94  &  0.97   & 0.98  &  0.98  &  0.98  &  0.99  &  0.99  &  0.99   & 0.99   & 0.99  & 1  & 1  \\
\hline \end{tabular}
\caption{Potential $\epsilon_n$-values the Quantiles, for matching support $\alpha, 0<\alpha <1.$
 }\label{table1}
\end{table}

\subsection{ABC with $d_k$ and a Euclidean distance}
\label{sec:ABC comparison}


\quad The goal is to compare simulated approximate posteriors  of  parametric ABC and  nonparametric ABC with $d_K.$  An ABC example   in Tavar\'{e} (2019, Lectures at Columbia University, \# 2, ``A Normal example'',  p. 35) is revisited.
$X_1,\ldots, X_n$ are $i.i.d.$ normal random variables, ${\cal N}(\theta, \sigma^2).$  The prior
for $\theta$ is uniform  $U(a,b)$ with $a\rightarrow -\infty$ and $b \rightarrow \infty.$ 
Attention is restricted
to the sample mean, ${\bar X}_n,$ since it  is  sufficient statistic. For fixed $a, b$ the posterior 
$\pi (\theta|{\bar X}_n)$ is  ${\cal N}(\theta, \frac{\sigma^2}{n})$  truncated in $(a,b).$
For the ABC-simulations  and a given $\epsilon^*$  it is assumed the observed ${\bar x}_n=0,$ $\theta^*$ is observed from $U(a,b)$ and is selected
when $\rho({\bar x}_n^*, {\bar x}_n=0)=|{\bar x}_n^*|\le \epsilon^*; | \cdot |$ is
 absolute value. A flat, ``0-1'', kernel is used to select $\theta^*.$

Approximate posterior densities appear in Figure 1 for  nonparametric ABC with $d_K$ and  parametric ABC with $| \cdot |.$
The Gaussian kernel is used by default in $R.$
The observed sample ${\bf X}=(X_1,\ldots, X_n)$ is from 
    ${\cal N}(0, 1),  \ n=100.$ 
 For the parametric ABC, given tolerance $\epsilon^*$   the steps in Tavar\'{e} (2019) are followed,
 using  ${\bar x}_n=0$   independently of the observed ${\bar x}_n.$

For nonparametric ABC with $d_K, \hat F_{\bf x}$ is used
 and $\epsilon$ is 
such that the number of selected $\theta^*$ 
from $U(-1,1)$ does not differ much from that of the parametric ABC. 
Randomness remains in the simulations but the number $N^*$  of drawn $\theta^*$ is large, $N^*=1,000,$  such that the number of $\theta^*$  selected ($N$ in Figure 1)  is also large enough for determining the
approximate  posterior. ${\bf X}_i^*$ is obtained from ${\cal N}(\theta_i^*,1)$  and $\theta_i^*$ is
selected if $d_K(\hat F_{\bf x},\hat  F_{{\bf x}_i^*})\le \epsilon, \ i=1,\ldots, N^*.$  The process is repeated for four values
of $(\epsilon, \epsilon^*).$  In Table 2, for the selected $\theta^*$ their mean $\bar {\theta^*},$  variance  
 and the mean square error of $\bar {\theta^*}$ from the mean $\theta=0$  of the posterior are calculated. 
When $\epsilon=.45$ and $\epsilon^*=1,$ at least 95\% of drawn $\theta^*$ are selected.  
\begin{table}[h!]
\begin{tabular}{|c|c|c|c|c|c|c|c| } \hline
 \multicolumn{8}{|c|}
{\bf Concentration:   Nonparametric ABC with $\hat F_{\bf x}$ and $d_K;$ Parametric ABC with ${\bar X}_n$} \\
\hline 
$\epsilon$-nonpar  & Mean $\theta_{select}^*$ & Var $\theta_{select}^*$  &MSE $\theta_{select}^*$    &  $\epsilon^*$-par  & Mean $\theta_{select}^*$ & Var $\theta_{select}^*$ 
&MSE $\theta_{select}^*$      \\ 
\hline
0.12  &  0.00456 &  0.022   &  0.022                  &                0.1  &  0.01820 &  0.0154  & 0.0157    \\ 
 \hline 
0.25 &  0.01780&  0.119 &  0.119      &     0.5  & -0.00816 &  0.0923 & 0.0924 \\    
\hline  
0.30 &  0.00774  & 0.192  & 0.192     &      0.6 & -0.01440 &  0.1340 &  0.1340 \\  
\hline
 0.45  & 0.01990  & 0.332 &  0.332    &       1.0 &  0.01550 &  0.3130 &  0.3130
 \\
\hline \end{tabular}
\caption{Mean, Variance and MSE of  $\theta^*_{select}$ 
 }\label{table2}
\end{table}

The MSE of parametric ABC posterior improves {\em uniformly in $\epsilon$} the nonparametric ABC.

\subsection{Comparison 
of parametric ABC with  F-ABC 
}
\label{s:FABC beats ABC}


\quad  The goal is to compare in simulations  parametric  ABC with {\em the least favorable} for concentration  F-ABC, 
{\em i.e.},  neglecting the additional concentration due to  {\bf 3)}{\em c)} of the F-ABC {\em Algorithm}.
Remark \ref{r:ABCcompFABC} is followed.  
Start  ABC with $d_K$ and $\epsilon$ and  for the selected $\theta_i^*$ in ABC,   draw  $M$ additional ${\bf x}^*$  to compute  $p_{match}(\theta_i^*).$
The  F-ABC posterior  for these  selected $\theta^*$ is obtained. 
For the non-selected  $\theta^*$ in ABC, $M$ additional ${\bf x}^*$ are drawn to compute the corresponding   $p_{match}(\theta^*).$  The  F-ABC posterior  for all $\theta^*$  drawn is then obtained.

In the simulations, very frequently,  the concentration (MSE) of the nonparametric  F-ABC  improves that of 
parametric ABC. In Tables 3 and 4 and the corresponding Figures 2 and 3,  examples are presented where the MSE of each method dominates the other.   The set-up  in section \ref{sec:ABC comparison} is used: $\epsilon^*= .15, \epsilon=.12, n=200, \theta=0, a=-1, b=1$ and   $N^*=1,000.$  For F-ABC,  $M=200$ ${\bf X}^*$-samples of size $n$ are drawn for each selected $\theta^*, $
but also for non-selected  $\theta^*.$  A flat, ``0-1'', kernel is used for selected $\theta^*$ in parametric ABC.

In Figures 2 and 3,  density plots with  Gaussian kernel and  corresponding histograms  are presented for ABC 
and
F-ABC. For the F-ABC approximate posteriors, the bandwidth was set at 0.05.
 Nonparametric F-ABC for selected $\theta^*$  is satisfactory compared with parametric ABC. F-ABC for all $\theta^*$ seems 
satisfactory for non-believers of $\theta^*$-exclusion with limited ${\bf x}^*$-data.
\begin{table}[h!]
\begin{tabular}{|c|c|c|c|c| } \hline
\multicolumn{5}{|c|}
{\bf Concentration:   Non Parametric ABC, F-ABC selected/drawn-Parametric ABC} \\
\hline 
\multicolumn {4}{|c|}{Nonparametric , $\epsilon=.12$} &   Parametric, $\epsilon^*=.15$ \\
\hline
Parameter  &  ABC   & F-ABC selected $\theta^*$   & F-ABC   all drawn $\theta^*$   & ABC   \\ 
\hline
Mean $\theta_{select}^*$  &  - 0.0916   &  -0.0865 &  -0.0859 &  -0.0117       \\ 
 \hline 
Variance $\theta_{select}^*$   &   0.0182  &   0.0105  &  0.0274  &  0.0107   \\    
\hline  
MSE $\theta_{select}^*$  &  0.0266  &   0.018 &    0.0348  &   0.0108  \\  
\hline
\end{tabular}
\caption{Mean, Variance and MSE of  $\theta^*_{select}$ 
 }\label{table3}
\end{table}
\begin{table}[h!]
\begin{tabular}{|c|c|c|c|c| } \hline
\multicolumn{5}{|c|}
{\bf Concentration:   Non Parametric ABC, F-ABC selected/drawn-Parametric ABC} \\
\hline 
\multicolumn {4}{|c|}{Nonparametric , $\epsilon=.12$} &   Parametric, $\epsilon^*=.15$ \\
\hline
Parameter  &  ABC   & F-ABC selected $\theta^*$  & F-ABC   all drawn $\theta^8$  & ABC   \\ 
\hline
Mean $\theta_{select}^*$  &  -0.00198 &  -0.00185 &  -0.00617 & 0.0112        \\ 
 \hline 
Variance $\theta_{select}^*$   &   0.0187  & 0.0111 & 0.0242 & 0.0138    \\    
\hline  
MSE $\theta_{select}^*$  &  0.0187 &  0.0111  & 0.0243  & 0.0139   \\  
\hline
\end{tabular}
\caption{Mean, Variance and MSE of  $\theta^*_{select}$ 
 }\label{table4}
\end{table}
\mbox{ }\\
\hspace*{1ex}  To compare the MSE improvement with F-ABC for selected $\theta^*,$
$K=1,000$ MSE comparisons\footnote{Used for higher accuracy. No need to be repeated.} are made
and the total number of  times, $T,$  F-ABC improves ABC is recorded.
The  parameters are $\epsilon=.12, \epsilon^*=.15, n=100, \theta=0, a=-1,b=1,N^*=100,  M=100.$ 
The process is repeated 50 times out of which 48  times $T>500,$ {\em i.e.}   F-ABC
for selected $\theta^*$  improves the MSE of parametric ABC.
A  histogram of the results appear in Table 4.
 To realize  50 comparisons, the process was repeated 55 times because   
of 5 non-terminations since  in F-ABC with $d_K$ 
there were simulations with no $\hat F_{{\bf x}^*}$ within  $\epsilon$  from $\hat F_{\bf x}.$
However,  in the majority of the remaining cases the number of ${\bf x}^*$   with F-ABC within $\epsilon$ 
from ${\bf x}$   exceeded that 
of  ABC. 


\subsection{ABC and F-ABC for all $\theta^*$  in $R^2$ with $d_K$ and half-spaces}
\label{subsec:F-ABC and half spaces}
\quad  ABC and F-ABC for all,  are implemented 
when    ${\bf X}=(X_1,\ldots,X_n) \in R^{nx2},$
with  $d_K$ used  for ${\bf X}^*$-
matching over  all 1-dimensional projections of ${\bf X}$ and ${\bf X}^*,$ 
or equivalently in  half-spaces,
 as explained 
in section \ref{sec:study epsilon}
 for the  sufficient, empirical measures $\mu_{\bf X}, \mu_{{\bf X}^*}.$ 

For $a, y \in R^2,$  $<a, y>$ is  the inner product  of $y$ and  $a,$
$|| \cdot ||$ is  Euclidean distance in $R^2.$
Using the notation in section 2,  $S({\bf X})=\mu_{\bf X}$  and $a \cdot {\bf X}=(<a,X_1>,\ldots, <a,X_n>) \in R^n,$
 $$\tilde \rho_n(\mu_{\bf X}, \mu_{{\bf X}^*})=
 \max_{a \in \{a_1,\ldots, a_{k_n}\} \subset  U_2}d_K(\hat F_{a \cdot {\bf X}}, \hat F_{a \cdot {\bf X}^*});
$$
 $a_1, \ldots, a_{k_n}$ are  are $i.i.d.$   uniform random vectors in $U_2=\{u=(u_1, u_2) \in R^2: ||u||=1\},$
 independent of ${\bf X}$ and ${\bf X}^*.$  Direction $a$ used in $\tilde \rho_n$ 
 has form $(\cos (\phi), \sin(\phi)),$ with  $\phi$ uniform  in $[0,\pi).$ 
$\tilde \rho_n$ approximates $\tilde \rho$ in (\ref{eq:empdist}) when $k_n \uparrow \infty,$ but a moderately large $k_n=k$ is adequate. For ABC and F-ABC the number of ${\bf X}^*$ $\epsilon$-matching ${\bf X}$   will decrease  as $k$ increases.

A sample ${\bf x}$ of size $n=50$ is observed from a bivariate normal  with means $\theta=(0,2),$ variances $1$ and covariance $.5.$   Assume
the parameter space is ${\bf \Theta}=[-1,2]x[-2,3] \subset R^2.$
Instead of drawing $\theta^*$ randomly from ${\bf \Theta},$  a discretization 
${\bf \Theta}^*$ of ${\bf \Theta}$
is used in order to observe the weights $p_{match(\theta^*)}$ along ${\bf \Theta}.$
With 
$NS=15$ equidistant $\theta_1^*$  and $\theta_2^*,$ respectively, in $[-1,2]$ and $[-2,3],$  obtain
$\theta^*=(\theta_1^*, \theta_2^*)$  in ${\bf \Theta^*}, 
 N^*=card({\bf \Theta^*})=225.$
Following Remark \ref{r:ABCcompFABC}, to obtain  $\tilde \rho_n$-ABC and  $\tilde \rho_n$-F-ABC posteriors,  one sample ${\bf X}^*$ is drawn initially
for each $\theta^*$ in ${\bf \Theta^*}.$ 
50 $a$-directions are used in $\tilde \rho_n,$  $\epsilon=.33$ and 
21 ${\bf X}^*$ match ${\bf X},$ 
thus selecting
21 $\theta^*$ 
from ${\bf \Theta^*}.$  With  F-ABC for all $\theta^* \in {\bf \Theta^*},$ without using {\bf 3}{\em c)} in the F-ABC Algorithm,    
$M=200$  independent  copies of   ${\bf X}^*$ are obtained for each  $\theta^* \in {\bf \Theta^*}.$
For the same  
50 $a$-directions and the $M+1$ matchings,
$p_{match}(\theta^*)$ in (\ref{eq:moregeneralthantopol})  is  calculated for  $\rho=\tilde \rho_n$  and 
$\epsilon=.33.$

In Figure 5, the ABC-posterior density  and the F-ABC for all $\theta^*$ posterior  histogram and density appear, created 
with $R$-functions $persp, \ hist3D$ and $persp3D,$ respectively.
Comparison of the ABC and  F-ABC densities   indicates higher concentration in the latter near the means $(0,2).$ 
Outside an area of (0,2), the $z$-values of the densities and the histogram  are 0 in all plots.
In ABC (all green),  the density's shape and the 0-values in the $z$-axis  are  due to  the  bivariate normal kernel used by default 
in $R$-function $kde2d$ needed in $persp.$ 
In F-ABC for all,  no kernel is used: the matching propostions, $p_{match}(\theta^*),$ are the weights,  frequencies and percentages, that   provide the 0's  and    nearly 0-values in the $z$-axis.
$ hist3D$ and $persp3D$ cannot be used in  F-ABC for the  selected  
$\theta^*.$ 

An additional  Example is  included
for the  Editors, AE and referees. New parameters are:  
$NS=10, card({\bf \Theta^*})=100, k=10, M=50.$ There are 11 selected $\theta^*$ with results  in Figure 6.
The small number of selected $\theta^*$  in both Examples  indicates the ABC-weakness with the choice of $\epsilon$-value,  which leads to repeated simulations for various $\epsilon$
{\em until} a ``satisfactory'' posterior is  obtained. F-ABC for all $\theta^*$ does not face this problem, 
{\em reducing} $\epsilon$'s influence.




 \section{Differences of  F-ABC and   ABC methods}
\label{sec:pmatch increases}

\quad  Main differences, some to appear in section \ref{sec:study epsilon},  are: the universal sufficient statistics, $\hat F_{\bf X}$ and $\mu_{\bf X},$ 
and matching via
$d_K;$ 
the F-ABC posterior 
for all $\theta^*$ drawn or used;
 the study and choice of 
$\epsilon;$ 
 the use of  $M$   ${\bf x}^*$  for each $\theta^* $
to obtain  $p_{match}(\theta^*),$ which is  the  $\theta^*$-weight and often depends on $\theta.$


For the last difference, in several models   it was observed  for $\tilde d, \rho$ generic distances that: 
\beeq
\label{eq:{eq:conditionMMSEP}0}
when \hspace{5ex}  \  d_{\bf \Theta}(\theta_1^*, \theta) \le d_{\bf \Theta}(\theta_2^*, \theta)  \Rightarrow  \tilde d(F_{\theta_1^*}, F_{\theta}) \le 
\tilde  d(F_{\theta_2^*}, F_{\theta})
\eneq
\beeq
\label{eq:conditionMMSEP} 
 \Rightarrow \ \forall  \   \epsilon>0, \hspace{3ex}
P_{\theta_1^*}[\rho(S({\bf X}^*), T(F_{\theta})) \le \epsilon] \ge  P_{\theta_2^*}[\rho(S({\bf X}^*), T(F_{\theta}) )\le \epsilon] .
\eneq


Implication (\ref{eq:{eq:conditionMMSEP}0})
usually holds.
 In  F-ABC
with $\tilde d= \rho=d_K, T(F_{\theta})=F_{\theta},  S({\bf X^*})=\hat F_{{\bf X}^*},$ 
 when (\ref{eq:conditionMMSEP}) holds  it will also hold, at least for large $n,$ when $F_{\theta}$ is replaced by $\hat F_{\bf X}.$
 For  families of c.d.fs in $R$ with densities $f_{\theta}$  such that $f_{\theta_1^*}(x)-f_{\theta_2^*}(x)$  changes sign once, 
 the upper probability of the last
implication in (\ref{eq:conditionMMSEP})  increases  to 1 with $n$ if 
$\theta^*$ gets closer to $\theta$ 
(Yatracos, 2020,  Propositions 7.2, 7.4  and Remark 7.2). An inequality similar to (\ref{eq:conditionMMSEP}) holds  for the  lower  bounds of these  probabilities (Proposition \ref{p:upper bounds order}). 
Thus, it is expected the F-ABC approximate posterior concentrates near  $\theta$ more than the ABC-posterior, as observed in the 
simulations in subsections \ref{s:FABC beats ABC} and \ref{subsec:F-ABC and half spaces}.
\bel
\label{l:why match prob 1}
The implications leading to (\ref{eq:conditionMMSEP}) hold for $i.i.d.$ normal random variables with mean $\theta$ and variance 1, 
$d_{\bf \Theta}=\rho=|.|, \ \tilde d=d_K, S({\bf X})=\bar X_n, \ T(F_{\theta})=\theta.$
\enl

For another difference, ${\bf X}$ is used without loss of generality instead of $S({\bf X})$ and measures 
$\nu, \mu$ are Lebesgue measures, each in a Euclidean space. For a function $h(\theta), \theta \in {\bf \Theta},$
 one goal is calculation  of  
\beeq
\label{eq:interest1}
E[h(\Theta)|{\bf X}={\bf x}]=\int_{\bf \Theta}h(\theta) \pi(\theta|{\bf x})d\theta.
\eneq
In ABC, (\ref{eq:interest1}) is
approximated using  the selected $\theta^*$  in ${\bf \Theta}_{ABC}^*,$
\beeq
\label{eq:h-ABC}
 \int_{\bf \Theta} h(\theta) \pi(\theta|{\bf x}) d\theta \approx
\int_{\bf \Theta} h(\theta) \pi(\theta)  \int f({\bf x}^*| \theta) K(\frac{{\bf x}^*-{\bf x}}{\epsilon_n}) d{\bf x}^* d\theta
\approx \sum_{{\bf \Theta}_{ABC}^*} h(\theta^*) \Pi_{ABC}(\theta^*|{\bf x});
\eneq
$\Pi_{ABC}(\theta^*|{\bf x})$ depends on $\pi(\theta), \epsilon_n, K$ and $f({\bf x}^*| \theta)$ which is usually intractable or unknown.

In F-ABC,  (\ref{eq:interest1}) is approximated using ${\bf \Theta}_n^*$  in (\ref{eq:selectedn}),
\beeq 
 \int_{\bf \Theta} h(\theta) \pi(\theta|{\bf x}) d\theta \approx \sum_{i=1}^{N} h(\theta^*_{sel,i})  p(\theta^*_{sel,i});
 \eneq
 $ p(\theta^*_{sel,i})$ depends  on additional ${\bf x}^*$ drawn, Kernel (\ref{eq:moregeneralthantopol}) with $\rho=d_K$  
and $\epsilon_n, i=1,\ldots, N.$


 
 \section{The Matching tools:
$ \hat F_{\bf X}, \mu_{\bf X}, d_K, \tilde \rho,  \tilde \rho_n, \epsilon \mbox{ and } \alpha$ } 
\label{sec:study epsilon}



{\quad \em Sufficiency, $\hat F_{\bf X}, \mu _{\bf X}, d_K, \tilde \rho, \tilde \rho_n$}
 

In ABC,
matching with  sufficient 
$S$  is  preferred since
$\pi(\theta | {\bf x})=\pi(\theta|S({\bf x})).$
When ${\bf X} \in R^{nx1}, \hat F_{\bf X}$ is sufficient 
being equivalent to the order statistic.  When ${\bf X}=(X_1,\ldots, X_n) \in R^{nxd}, d>1,$  
and $X_1, \ldots, X_n$ are either $i.i.d.$  or exchangeable,
the empirical measure, $\mu_{\bf  X},$
\beeq
\label{eq:mu}
\mu_{\bf X}(A)=n^{-1}\sum_{i=1}^{n}I_A(X_i), \ A \in {\cal B}_d,
\eneq
is sufficient, respectively by,  Dudley (1984,  Theorem 10.1.3, p. 95)  and
de Finetti's Theorem,  {\em e.g.},  Lauritzen (2007, in Statistical Implications section);
 ${\cal B}_d$ are the Borel sets in $R^d.$ 
When $d>1,$ for some models  $\hat F_{\bf X}$ may be nearly sufficient but still better than  guessing $S.$

For ${\bf x} \in R^d, d>1,$ to guarantee sufficiency, $\mu_{\bf X}$ is  used for $\epsilon$-matching ${\bf X}$ with ${\bf X}^*.$   
As explained below, instead of using for matching the usual form of Total Variation distance,
\beeq
\label{eq:empL1}
\tilde \rho(\mu_{\bf X},\mu_{{\bf X}^*})=\sup_{A \in {\cal B}_d}|\mu_{\bf X}(A)-\mu_{{\bf X}^*}(A)|=TV(\mu_{\bf X},\mu_{{\bf X}^*}),
\eneq
 the supremum in (\ref{eq:empL1}) is over all  half-spaces, 
\beeq
\label{eq:half-space}
A(a, t)=\{y \in R^d: <a, y> \le t\}, \ t \in R, a \in  U_d=\{u=(u_1,\ldots, u_d) \in R^d: ||u||=1\};
\eneq
$<a, y>$ is the inner product of $y$ and  $a, || \cdot ||$ is  Euclidean distance in $R^d.$ Then, 
\beeq
\label{eq:empdist}
\tilde \rho(\mu_{\bf X}, \mu_{{\bf X}^*})= \sup_{a \in U_d}  \sup_{t \in R} |\mu_{\bf X}(A(a,t))-\mu_{{\bf X}^*}(A(a,t))|
= \sup_{a \in U_d} d_K(\hat F_{a \cdot {\bf X}}, \hat F_{a \cdot {\bf X}^*}),
\eneq
$a \cdot {\bf X}=(<a,X_1>,\ldots, <a,X_n>) \in R^n.$

 In practice, $\tilde \rho(\mu_{\bf X}, \mu_{{\bf X}^*})$ is approximated by
\beeq
\label{eq:empdistapprox}
\tilde \rho_n(\mu_{\bf X}, \mu_{{\bf X}^*})=
 \max_{a \in \{a_1,\ldots, a_{k_n}\} \subset  U_d} \sup_{t \in R} |\mu_{\bf X}(A(a,t))-\mu_{{\bf X}^*}(A(a,t))|= \max_{a \in \{a_1,\ldots, a_{k_n}\} \subset  U_d}d_K(\hat F_{a \cdot {\bf X}}, \hat F_{a \cdot {\bf X}^*}).
\eneq
where $a_1, \ldots, a_{k_n}$ are either a discretization of $U_d$ or  $i.i.d.$   uniform in $U_d,$ independent of ${\bf X}$ and ${\bf X}^*,$ leading to approximate sufficiency. Using $A=A(a,t)$ in  (\ref{eq:mu}),
\beeq
\label{eq:back to cdf}
I_{A(a,t)}(X_i)=1 \iff <a,X_i> \le t  \Rightarrow \mu_{\bf X}(A(a,t))=\frac{Card( <a, X_i> \le t, \ i=1,\ldots,n)}{n}
=\hat F_{a \cdot {\bf X}}(t),
\eneq
and the last equalities in (\ref{eq:empdist}) and  (\ref{eq:empdistapprox}) follow,
relating  $\tilde \rho$ over all  half-spaces with $d_K$-distance over  all 1-dimensional projections of ${\bf X}, {\bf X}^*.$
Hence, in applications, 
${\bf X}$ will match  ${\bf X}^*$
when the last term in (\ref{eq:empdistapprox}) is less than or equal to $\epsilon_n,$  with the $R$-functions used for $d=1.$

{\em$\tilde \rho$ and $\tilde \rho_n$}

 If $P$ and $Q$ are probabilities in $(R^d, {\cal B}_d)$ which are
 equal over all half-spaces, $A(a,t),$ in (\ref{eq:half-space}),  then $P$ and   $Q$  are equal for every 
$A \in {\cal B}_d$ (Cram\'er and Wold, 1936).
When ${\bf X}$'s coordinates follow the unknown probability  $P \in {\cal P}$  and $\tilde \rho$ is defined in (\ref{eq:empdist}),  Beran and Millar (1986, p. 431-433, Theorem 3, p. 436) obtained  confidence sets $\{Q \in {\cal P}: \tilde \rho(\mu_{\bf X},Q)<c\}$  for
$P$ using  $\tilde \rho_n$ with $a_1,\ldots, a_{k_n} \ i.i.d.$  uniform on $U_d,$  and  showed that when $k_n \uparrow \infty$ as $n \uparrow \infty,$ then $\lim_{n\rightarrow \infty} \tilde \rho_n(P,Q)=\tilde \rho(P,Q)$ with probability 1  and asymptotically the 
required  
coverage is achieved.
 
{\em Pertinent properties of  $\hat F_{\bf X}, d_K, \mu_{\bf X}$}

$\hat F_{\bf X}$ and  $d_K$ satisfy desired properties for summary statistics (Fearnhead and Prangle, 2012, Frazier {\em et al.}, 2018) when $F_{\theta}$  is the parameter of interest: {\em  a)} $ E\hat F_{\bf X}=F_{\theta},  b) F_{\theta_1}=F_{\theta_2}$ 
implies $\theta_1=\theta_2$ due to identifiability,
and {\em c)} there are various types of $\hat F_{\bf x}$ 's convergence to $F_{\theta},$ including $d_K$-convergence. When  $T(F_{\theta})=\theta$  and 
$T$ is continuous with respect to $d_K$ and a metric $d_{\bf \Theta}$ on ${\bf \Theta},$  
it is expected that $T(\hat F_{\bf X})$ as estimate of $\theta$ will inherit  convergence properties  of $\hat F_{\bf X}$ to $F_{\theta}.$ Similar results hold for the empirical measure, $\mu_{\bf X}$, its  corresponding probability $P_{\theta}$ 
and the class of half-spaces which is Vapnik-Cervonenkis class of sets with index (d+1), see, {\em e.g.}  Dudley (1978).

$d_K(\hat F_{{\bf x}^*}, \hat F_{\bf x})$ is not continuous function in $R^n$ at ${\bf x}$ 
since it cannot be 
 smaller than $\frac{1}{n}$  for {\em all} ${\bf x}^*$ at 
Euclidean distance $\delta>0$ from ${\bf x}.$
This makes $d_K$ different from other $\rho$-distances used in ABC, (\ref{eq:topol}),  (\ref{eq:topolS}); see, {\em e.g.} Bernton {\em et al.} (2019, p. 39, proof of Proposition 3.1).

\bel
\label{l:Kolmandempiricalproperty}
For any observed samples of size $n,$  ${\bf x}^* \neq {\bf x}_{\sigma(1:n)}  \in R^d, d\ge 1,$
\beeq
\label{eq:Kolmandempiricalproperty1}
d_K(\hat F_{\bf x}, \hat F_{{\bf x}^*}) \ge \frac{1}{n};
\eneq
${\bf x}_{\sigma(1:n)}$ denotes a vector, permutation of the ${\bf x}$ components.
Thus,
\beeq
\label{eq:Kolmandempiricalproperty2}
d_K(\hat F_{\bf x}, \hat F_{{\bf x}^*})=0 \iff {\bf x}^*={\bf x}_{\sigma(1:n)}.
\eneq
\enl

{\em $\epsilon_n, \alpha$ and $d_K$}

For matching support probability  $\alpha$  in (\ref{eq:tolfabcforone}),
the F-ABC tolerance $\epsilon_n$  satisfies 
 \begin{equation}
\label{eq:criterion1}
P[d_K(\hat F_{{\bf X}^*}, \hat F_{{\bf X}}) > \epsilon_n] = 1- \alpha, \ 0\le \alpha \le 1. 
\end{equation}
An upper bound $\epsilon_{n,B}$ on $\epsilon_n$ is obtained
equating an upper probability bound in (\ref{eq:criterion1})
with $1-\alpha;$
see Lemma \ref{l:upperepsilon}. 
Conditionally on ${\bf X}={\bf x},$ 
$\epsilon_{n,B}({\bf x})$  is similarly  obtained 
under
$F_{\theta^*}.$ 
The $\epsilon_n$ upper bounds follow for  ${\bf X}$ and ${\bf X}^*  \in R^{nxd}, d=1.$ When $d>1,$  similar results hold presented 
after the Proof of Proposition  \ref{p:fiducmatchtolunc}.


  \bep 
\label{p:fiducmatchtolunc}
 Let ${\bf X}$ be a sample of  $n$ random variables from  cumulative distribution $F_{\theta},$  with $\theta$ unknown,  let ${\bf X}^*$ be a simulated $n$-size sample from a sampler used for 
$\theta^*$
and  let  $\alpha$ be the 
matching support probability for the 
tolerance $\epsilon_n$ in   (\ref{eq:criterion1}); $  0\le \alpha < 1.$\\
a) 
The upper bound for $\epsilon_n$ is 
\begin{equation}\label{eq:fidutol}
\epsilon_{n,B}(\theta, \theta^*)=
d_K(F_{\theta}, F_{\theta^*}) +\sqrt{\frac{2}{n} \ln \frac{4} {1-\alpha}}\ge \sqrt{\frac{2}{n} \ln 4} .
\end{equation}
b) Conditionally on ${\bf X}={\bf x},$ the upper bound for $\epsilon_n$ is
\begin{equation}\label{eq:fidutolx}
\epsilon_{n,B}({\bf x},\theta^*)=d_K(\hat F_{\bf x},  F_{\theta^*})+\sqrt{\frac{1}{2n}\ln \frac{2}{1-\alpha}}\ge \delta_n({\bf x},\theta^*)
+ \sqrt{\frac{1}{2n}\ln 2}.
\end{equation}
In practice, $\min\{\epsilon_{n,B}(\theta, \theta^*),1\}$ and $\min\{\epsilon_{n,B}({\bf x}, \theta^*),1\}$ are used.

\enp

(\ref{eq:fidutol}) and  (\ref{eq:fidutolx} provide a structure for the tolerance.   Since $F_{\theta}$ is unknown and $\theta^* \in {\bf \Theta^*},$ uniform upper bounds are useful. 
Since   $\hat F_{{\bf X}}$ is with high probability at $d_K$-distance $\frac{C_n}{\sqrt{n}}$
from $F_{\theta},$    a  plausible  choice for the uniform upper bounds  
of  $d_K(F_{\theta}, F_{\theta^*})$   and $d_K(\hat F_{\bf x},  F_{\theta^*})$ 
is  $\frac{C_n^*}{\sqrt{n}},$ with  $C^*_n>C_n>0.$ 
Probability bounds are rarely  tight and,  in practice, $\epsilon_n$ is determined via  simulations;
see Table 1 in  subsection \ref{s:epsilonand1-alpha}.




The next Proposition indicates that for the lower  bounds  on the Probabilities in 
(\ref{eq:conditionMMSEP}),  the same inequality holds 
when $d_K(F_{\theta_1^*}, F_{\theta}) < d_K(F_{\theta_2^*}, F_{\theta}) <\epsilon_n.$

\bep
\label{p:upper bounds order}
For $n$  i.i.d. random vectors in $R^d$  with c.d.f. $F_{\theta^*}$ and $n$ large:
\beeq
\label{eq:upper bounds order}
P_{\theta^*}[d_K(F_{{\bf X}^*}, \hat F_{\bf X}) \le \epsilon_n]\ge 1- C_1^*(d)  \cdot \exp \{-n\cdot C_2^*(d) \cdot (\epsilon_n-d_K(F_{\theta^*},F_{\theta}))^2\};
\eneq
$C_1^*(d), \ C_2^*(d)$ are positive constants.
\enp

 
\section{Asymptotics}
\label{sec:asy}

\quad  Results obtained  for Kolmogorov distance, $d_K,$ when ${\bf X} \in R^{nxd},$ hold also for the stronger distance 
(\ref{eq:empdist}) using $d_K$ on all half-spaces in $R^d.$

In ABC, one question of interest is
whether $\pi_{abc}(\theta|B_{\epsilon})$
 converges to $\pi(\theta|{\bf x})$ when   ${\bf x}$ stays fixed and $\epsilon=\delta_m \downarrow 0$ 
 as $m$ increases.




\bep
\label{p:conveoposterior} 
Use  the notation in section 2, for {\em ABC} and {\em F-ABC}  with $S({\bf X})=\hat F_{\bf X}, \rho=d_K, $  $n$ fixed 
and $B_{\epsilon_n}$ in (\ref{eq:B}).
Under the exchangeability assumption, i.e.
$f({\bf y}|\theta)=f({\bf y}_{\sigma(1:n)}|\theta)$  for any permutation
${\bf y}_{\sigma(1:n)}$ of ${\bf y},$ and 
 with    $\delta_m \downarrow 0$ as $m$ increases, 
\beeq
\label{eq:conveoposterior}
\lim_{m \rightarrow \infty} \pi_{u}(\theta|B_{\delta_m})=\pi(\theta | {\bf x}), \hspace{5ex}  u=abc, {f\mbox{-}abc}.
 \eneq 
For continuous   ${\bf X},$ $({\cal  Y},{\cal C}_{\cal Y})$ is $R^{nxd}$ with the Borel sets,  ${\cal B},$  and $\Theta$ takes values in $R^k, k \le d.$
\enp

Another  question of interest for ABC  is whether
 the posterior  
$\pi_{abc}(\theta|B_{\epsilon_n})$ will place
increasing probability mass around $\theta$ as $n$ increases to infinity (Fearnhead, 2018), {\em i.e.} Bayesian consistency.
Posterior concentration is proved for ABC and F-ABC,
 initially  for fixed size $\zeta$-neighborhood
when $T(F_{\theta})$ is the quantity of interest;
$T$ is a functional, $\zeta>0.$ 

\bep 
\label{p:concentration}
Use the notation in section 2 and  let 
 ${\cal F}_{\bf \Theta}=\{F_{\theta}, \theta \in {\bf \Theta} \}$ be  subset of a metric space $({\cal F}, d_{\cal F})$ of c.d.fs. Assume\\
a) $d_{\cal F} (\hat F_{\bf X}, F_{\theta}) \le \frac{o(k_n)}{k_n}, k_n \uparrow \infty$
and $P_{\theta}^{(n)}$-probability $\uparrow 1,$  as $n$ increases,
 and \\ 
b) $T$ is  a continuous functional on ${\cal F} $ with values in a metric space $({\cal T}, d_{\cal T}).$\\
 Then, 
for 
{\em  ABC} and {\em F-ABC},  $S({\bf X})=\hat F_{\bf X},  \rho=d_{\cal F}$ and   for any $\zeta>0$
\beeq
\label{eq:concfixedarea}
\lim_{n \rightarrow \infty} \Pi_{u}[\theta^*: d_{\cal T}(T(F_{\theta^*}),T(F_{\theta})) \le {\zeta} | B_{\epsilon_n}]=
1,  \hspace{5ex}  u=abc, {f\mbox{-}abc};
\eneq
\beeq
\label{eq:Bd_T}
B_{\epsilon_n}= \{  {\bf x}^*: d_{\cal F}(\hat F({\bf x}^*), \hat F({\bf x} )) \le \epsilon_n\}, \epsilon_n \downarrow 0
\mbox{ as } n \uparrow \infty.
\eneq
\enp

\ber
\label{r:applicationhatF} In Proposition \ref{p:concentration},
assumption a) 
holds for $d_{\cal F}=d_{K}, k_n=\sqrt{n};$  special case of interest  in b)  when $T(F_{\theta})=\theta$ and $d_{\cal T}=d_{\bf \Theta},$ 
the metric on ${\bf \Theta}.$
\enr

To confirm Bayesian consistency 
for shrinking $d_{\cal T}$-neighborhoods  of $T(F_{\theta}),$ let $w$ be
the modulus of continuity of $T,$ {\em i.e.}
\beeq
\label{eq:modcont}
w(\tilde \epsilon)=\sup\{d_{\cal T}(T(F_{\theta}), T(F_{\eta})): d_{\cal F}(F_{\theta}, F_{\eta}) \le \tilde \epsilon; \theta \in {\bf \Theta}, \eta \in {\bf \Theta}\}, \  \tilde \epsilon >0.
\eneq
Consistency was established for $\zeta$-$d_{\cal T}$-neighborhood of $T(F_{\theta})$  when 
(\ref{eq:entildee}) holds, {\em i.e.} when 
$$ 
\epsilon_n \le \tilde \epsilon -\frac{2o(k_n)}{k_n},
$$
 thus it holds for the  smallest  $\tilde \epsilon$-value,
\beeq
\label{eq:smallesttildeepsilon}
\tilde \epsilon=\epsilon_n+\frac{2o(k_n)}{k_n}
\eneq 
 and
since for $\zeta_n$-$d_{\cal T}$-neighborhood of $T(F_{\theta})$ 
$$\zeta_n=w(\tilde \epsilon)$$
it follows that
\beeq
\label{eq:invmodcont}
\zeta_n=w(\epsilon_n+\frac{2o(k_n)}{k_n}) \ge w(\frac{2o(k_n)}{k_n})  .
\eneq

\bel
\label{shrinkingneighrate}
Under the assumptions of Proposition \ref{p:concentration}, the shortest $d_{\cal T}$-shrinking neighborhood 
of $T(F_{\theta})$ for which Bayesian consistency holds has radius $w(\epsilon_n + \frac{2o(k_n)}{k_n}) \ge
w(\frac{2o(k_n)}{k_n})  .$
\enl

\ber
\label{r:conclusion1}
The rate of posterior concentration around $T(F_{\theta})$ depends, as expected,  on the rate in probability, $k_n^{-1},$ of the $d_{\cal F}$-concentration of
$T(\hat F_{\bf X})$ around $T(F_{\theta})$ which is not under the user's control, the tolerance $\epsilon_n$ and the modulus of continuity,
$w,$  of $T.$ Similar conclusions in a different set-up have been obtained by Frazier {\em et al.} (2018).
\enr

\section{Annex}

\quad {\bf Proof of Lemma \ref{l:why match prob 1}:}  The first implication holds from the corresponding models, {\em w.l.o.g.} for $\theta< \theta_1^*<\theta_2^*,$   by observing that $d_K(F_{\theta},F_{\theta_1^*})=F_{\theta}(.5(\theta+\theta_1^*))-
F_{\theta_1^*}(.5(\theta+\theta_1^*))$ and comparing  with $d_K(F_{\theta},F_{\theta_2^*}).$
The last implication holds from the assumption since 
$$G(\theta^*)=\Phi[\sqrt{n}(\epsilon+\theta-\theta^*)]-\Phi[\sqrt{n}(-\epsilon+\theta-\theta^*)]$$  is decreasing in $\theta^*$ when $\theta^*>\theta$ and 
increasing in $\theta^*$ when $\theta^*<\theta,$ and determines the probabilities
in (\ref{eq:conditionMMSEP})  for $\theta^*=\theta_1^*, \theta_2^*.$ Indeed,
$$P_{\theta^*}(|\bar X_n-\theta|\le \epsilon)=P_{\theta^*}(-\epsilon+\theta\le \bar X_n \le \epsilon+\theta)
=P[\sqrt{n}(-\epsilon+\theta-\theta^*)\le Z \le \sqrt{n}(\epsilon+\theta-\theta^*)]$$
$$=\Phi[\sqrt{n}(\epsilon+\theta-\theta^*)]-\Phi[\sqrt{n}(-\epsilon+\theta-\theta^*)]=G(\theta^*)$$
$$G'(\theta^*)=-\sqrt{n}\phi(\sqrt{n}(\epsilon+\theta-\theta^*)+\sqrt{n}\phi(\sqrt{n}(-\epsilon+\theta-\theta^*)<0$$
$$\iff \phi(\sqrt{n}(-\epsilon+\theta-\theta^*)) <\phi(\sqrt{n}(\epsilon+\theta-\theta^*)) \iff -(-\epsilon+\theta-\theta^*) ^2<-(\epsilon+\theta-\theta^*)^2$$
$$\iff 2 \epsilon(\theta-\theta^*)<-2 \epsilon(\theta-\theta^*)\iff  4 \epsilon(\theta-\theta^*)<0  \iff   \theta<\theta^*, $$
hence if $\theta <\theta^*, G(\theta^*)$ is decreasing in $\theta^*, \  \theta< \theta_1^*<\theta_2^* \Rightarrow G(\theta_1^*)>G(\theta_2^*).$
 For $\theta^* <\theta,  G(\theta^*)$ is increasing, $\theta_2^*<\theta_1^* <\theta \Rightarrow G(\theta_2^*)<G(\theta_1^*). \hspace{5ex}\Box$

 {\bf Proof of Lemma \ref{l:Kolmandempiricalproperty}:} The smaller $d_K$-distance between $\hat F_{\bf x}$ and $\hat F_{{\bf x}^*}$ occurs when  ${\bf x}, {\bf x}^*$   differ by a small $\delta >0$   in one coordinate  of one  observation and  their distance is $\frac{1}{n}. \hspace{5ex} \Box$

\bel
\label{l:upperepsilon}
Let ${\bf X}={\bf x}, {\bf X}^*={\bf x}^*$ and let  $U(n,\epsilon)$  be  positive function defined for 
positive integers $n$  and  $\epsilon>0, 0\le \alpha \le 1, $ such that
\beeq
\label{eq:upperepsilon}
1-\alpha=P[d_K(\hat F_{\bf x},\hat F_{{\bf x}^*})>\epsilon]\le U(n,\epsilon).
\eneq
Let $\epsilon_B: U(n,\epsilon_B)=1-\alpha.$ Then $\epsilon_B \ge \epsilon.$
\enl

{\bf Proof of Lemma \ref {l:upperepsilon}:} Since $U(n,\epsilon_B)=1-\alpha,$ 
$$P[d_K(\hat F_{\bf x},\hat F_{{\bf x}^*})> \epsilon_B]\le U(n,\epsilon_B)=1-\alpha =P[d_K(\hat F_{\bf x},\hat F_{{\bf x}^*})>\epsilon]$$ 
which implies $\epsilon_B \ge \epsilon.$ \hspace{5ex} $\Box$

\beth (Dvoretzky, Kiefer and Wolfowitz, 1956, and  Massart, 1990,  providing the tight constant)
\label{t:DKWM}
 Let $\hat F_{\bf Y}$ denote the empirical c.d.f of the size $n$ sample ${\bf Y}$  of i.i.d. random variables  obtained from cumulative distribution $F.$
Then, for any $\epsilon >0,$ 
\begin{equation}
\label{eq:DKWM}
P[d_K(\hat F_{\bf Y},F) > \epsilon]\le U_{DKWM}=2 e^{-2n\epsilon^2}
\end{equation}
\enth

{\bf Proof of Proposition  \ref{p:fiducmatchtolunc}:} {\em a)}
$$P[d_K(\hat F_{ {\bf X}^*},\hat F_{{\bf X}})>\epsilon_n] \le P[d_K(\hat F_{ {\bf X}^*}, F_{\theta^*})
+d_K(F_{\theta^*}, F_{\theta}) +d_K( F_{\theta}, \hat F_{ {\bf X}})>\epsilon_n]  $$
$$\le P[d_K(\hat F_{{\bf X}^*}, F_{\theta^*}) >\frac{\epsilon_n -d_K(F_{\theta^*}, F_{\theta})}{2}] 
+  P[d_K(\hat F_{  {\bf X}}, F_{\theta}) >\frac{\epsilon_n -d_K(F_{\theta^*}, F_{\theta})}{2}] $$
$$\le 4 \exp\{-\frac{n}{2}(\epsilon_n -d_K(F_{\theta^*}, F_{\theta}))^2\}$$
 The right side of the last inequality, obtained from 
(\ref{eq:DKWM})  is made equal to $1-\alpha,$ 
$$4 \exp\{-\frac{n}{2}(\epsilon_{n,B} -d_K(F_{\theta^*}, F_{\theta}))^2\}=1-a 
\iff  \epsilon_{n,B}=d_K(F_{\theta^*}, F_{\theta})+\sqrt{\frac{2}{n} \ln \frac{4}{1-\alpha}}.$$
{\em b)} $P[d_K(\hat F_{ {\bf X}^*},\hat F_{{\bf x}})>\epsilon_n] \le P[d_K(\hat F_{ {\bf X}^*}, F_{\theta^*})
+d_K(F_{\theta^*}, \hat F_{ {\bf x}})>\epsilon_n] \le  2 \exp {\{-2n(\epsilon_n-d_K(F_{\theta^*}, \hat F_{ {\bf x}}))^2\}} $\\
 obtaining with matching support probability $\alpha,$
$$\epsilon_{n,B}({\bf x})=d_K(F_{\theta^*}, \hat F_{ {\bf x}})+\sqrt{\frac{1}{2n}\ln \frac{2}{1-\alpha}}.  \hspace{5ex} \Box $$

Generalizations 
of (\ref{eq:DKWM}) in $R^d$  have been obtained, at least,  by Kiefer and Wolfowitz (1958), Kiefer (1961)
and Devroye (1977); $ d>1.$  The differences in  upper bound $U$  in (\ref{eq:DKWM}) are in the multiplicative constant,  in the exponent of the exponential and on the sample size
for which the exponential bound holds which may also depend on $\epsilon.$ 
The 
constants used
 are  not determined  except for  Devroye (1977).

For example, following the Proof in Proposition \ref{p:fiducmatchtolunc} {\em b)}, conditionally on ${\bf X}={\bf x}:$ \\ 
{\em i)}  Using Kiefer and Wolfowitz (1958),
with the upper bound in (\ref{eq:DKWM})  $U_{KW}=C_1(d)e^{-C_2(d)n \epsilon^2},$
$$\epsilon_{n,B}({\bf x}, \theta^*)=
d_K(\hat F_{\bf x},  F_{\theta^*}) +\sqrt{\frac{1}{nC_2(d)} \ln \frac{C_1(d)}{1-\alpha}}. $$
{\em ii)}  Using Kiefer (1961),
with the upper bound in (\ref{eq:DKWM})   $U_{K}=C_3(b, d)e^{-(2-b)n \epsilon^2},$ for every $ b \in (0,2),$
$$\epsilon_{n,B}({\bf x}, \theta^*)=d_K(\hat F_{\bf x},  F_{\theta^*})+\sqrt{\frac{1}{n(2-b)} \ln \frac{C_3(b, d)}{1-\alpha}}. $$
{\em iii)} Using Devroye (1977), with the upper bound in (\ref{eq:DKWM}) $U_{De}= 2e^2(2n)^de^{-2n\epsilon^2}$  valid for $n\epsilon^2 \ge d^2, $
$$\epsilon_{n,B}({\bf x}, \theta^*)=d_K(\hat F_{\bf x},  F_{\theta^*})+ \sqrt{\frac{1}{2n}[\ln \frac{2}{1-\alpha}+2+d \ln (2n)]}.$$

 \ber
\label{r:epsilondecomp}
In (\ref{eq:fidutol}),  (\ref{eq:fidutolx}) and in  i)-iii), $\epsilon_{n,B}$  is the sum of the  model discrepancy
of $F_{\theta^*}$ from either $F_{\theta}$ or $\hat F_{\bf x}$  and  a confidence 
 term, determined, respectively,  under  both 
$F_{\theta}$ and  $F_{\theta^*}$ or the latter only. 
$\epsilon_{n,B}$  is independent of $\theta^*$ and
${\bf x}$  in  (\ref{eq:fidutol}) only. In all cases, since $F_{\theta}, F_{\theta^*}$ are unknown, a bound  will be used
for   $d_K(F_{\theta}, F_{\theta^*}),  d_K(\hat F_{\bf x},  F_{\theta^*}).$
\enr

{\bf Proof of  Proposition \ref{p:upper bounds order}:} Follows along the first three  lines in the proof of  Proposition  \ref{p:fiducmatchtolunc} {\em a)},  with the exponential upper bound obtained using the $U_{KW}$ above
in {\em i)} (Kiefer and Wolfowitz, 1958), with $C_1^*(d), C_2^*(d)$ the adjustments of $C_1(d), C_2(d).$  \hspace{5ex} $\Box $

{\bf Proof of Proposition \ref{p:conveoposterior}:} The arguments used for ABC  hold for  F-ABC.\\
{\em a)} ${\cal Y}$ discrete:  The ABC posterior with $\rho=d_K$  in (\ref{def: f-abcposterior})  is 
 $$\pi_{abc}(\theta|B_{\delta_m})
 =\frac{\pi(\theta) \cdot
\int_{\cal Y}  I_{B_{\delta_m}}({\bf y}^*)  f({\bf y}^*|\theta) \mu(d{\bf y}^*)}
{\int_{\bf \Theta}  \pi(s)  \int_{{\cal Y}}I_{B_{\delta_m}}({\bf y}^*) f({\bf y}^*|s) \mu(d{\bf y}^*)  \ \nu(ds)    }.
  $$
With integral denoting sum,
 it is enough to prove that the integral in the numerator of  $ \pi_{abc}(\theta|B_{\delta_m})$
 is proportional
to $f({\bf x}|\theta).$
     
For  $A \in {\cal C}_{\cal Y},$    let 
$$Q_{\theta}(A)=\int_A f({\bf y}^*|\theta) \mu(d{\bf y}^*), \ A \in {\cal A}.$$
$Q_{\theta}$ is a probability measure on ${\cal C}_{\cal Y}. $

Since $n$ and ${\bf x}$  are fixed, for $\delta_k \ge \frac{1}{n}> \delta_{k+1}$
\beeq
\label{eq:decreasingBs}
B_{\delta_1}  \supseteq B_{\delta_2}  \supseteq \ldots    \supseteq B_{\delta_k}  
\eneq
and  from  Lemma \ref{l:Kolmandempiricalproperty} for $m >k,  B_{\delta_m}=\{{\bf{x}}_{\sigma(1:n)}\}.$ 
Therefore, 
\beeq
\label{eq:limitdecreasingBs}
 \lim_{m \rightarrow \infty} B_{\delta_m}=\cap_{m=1}^{\infty} B_{\delta_m}=\{{\bf{x}}_{\sigma(1:n)}\}
\eneq
and 
\beeq
\label{eq:lastdiscrete}
\lim_{m \rightarrow \infty} \int_{\cal Y}  I_{B_{\delta_m}}({\bf y}^*)  f({\bf y}^*|\theta) \mu(d{\bf y}^*)\
=\lim_{m \rightarrow \infty}  Q_{\theta}(B_{\delta m})=  Q_{\theta}(\cap_{m=1}^{\infty} B_{\delta_m})=
f({\bf {x}}|\theta) \mu(\{{\bf x}_{\sigma(1:n)}\}),
\eneq
with the last equality due to exchangeability of $f({\bf {x}}|\theta).$

{\em b)} ${\cal Y}$ continuous: Then, the  right side of (\ref{eq:lastdiscrete}) vanishes, since 
$\mu(\{{\bf x}_{\sigma(1:n)}\})=0.$ A different approach is used, via the notion
of regular conditional probability.

When ${\cal Y}$ is a Euclidean space $R^{nxd}$ with  Borel $\sigma$-field, ${\cal B}_d,$ 
and  ${\Theta}$ takes values in $R^k, k \le d,$   
the integral in the numerator of $\pi_{abc}(\theta|B_{\delta_m}),$ 
$$ \int_{\cal Y}  I_{B_{\delta_m}}({\bf y}^*)  f({\bf y}^*|\theta) \mu(d{\bf y}^*)$$  
is a  regular conditional probability,  $P[{\bf X}^* \in B|\Theta=\theta], B=B_{\delta_m}$  (Breiman, 1992, Chapter 4, p. 79, Theorem 4.34),  {\em i.e.}, with $\theta$ fixed, it is 
a probability  for $B \in {\cal B}_d$   and with fixed $B$ it is a version of 
the conditional density, $\theta \in {\bf \Theta}.$ Thus, for fixed ${\theta},$  from (\ref{eq:limitdecreasingBs}),
$$ \lim_{m \rightarrow \infty} P[{\bf X}^* \in B_{\delta_m}|\Theta=\theta]= P[\{{\bf x}_{\sigma(1:n)}\}|\Theta=\theta]$$
and due to exchangeability is proportional to $f({\bf x}|\theta)$ {\em a.s.} . $\hspace{5ex}\Box$

{\bf Proof of Proposition \ref{p:concentration}:}  The arguments used for ABC  hold for  F-ABC.\\
For the probability in (\ref{eq:concfixedarea}), using 
 (\ref{eq:fabcposteriorProb})  for ABC
 with   
\beeq
\label{eq:Hforconcfixedarea}
 H=\{\theta^*: d_{\cal T}(T(F_{\theta^*}),T(F_{\theta})) \le {\zeta}\},
\eneq
\beeq
\label{eq:ABCposteriorprob}
 \Pi_{abc}(H|B_{\epsilon_n})= \frac{  \int_{\bf \Theta}I_ {H} (\theta^*) \pi(\theta^*) \cdot
\int_{\cal Y}  I_{B_{\epsilon_n}}({\bf y}^*)  f({\bf y}^*|\theta^*) \mu(d{\bf y}^*) \nu(d\theta^*)}
{\int_{\bf \Theta}  \pi(s)  \int_{{\cal Y}}I_{B_{\epsilon_n}}({\bf y}^*) f({\bf y}^*|s) \mu(d{\bf y}^*)  \ \nu(ds)    }
=\frac{\int_{\bf \Theta}\pi(\theta^*) \cdot P_{\theta^*}^{(n)}(H\cap B_{\epsilon_n })\nu(d\theta)} {\int_{\bf \Theta}\pi(s) \cdot P_{s}^{(n)}(B_{\epsilon_n}) \nu(ds)}.
\eneq
$ P_{\theta^*}^{(n)}(H\cap B_{\epsilon_n })$ in the numerators of (\ref{eq:ABCposteriorprob}) will be bounded below using continuity of $T$  and triangular inequality.

Since $T$ is continuous, for $\zeta>0$ there is $\tilde \epsilon >0$ such that if 
$$
d_{{\cal F}}(F_{\theta^*}, F_{\theta}) \le \tilde \epsilon \mbox{ then }
 d_{\cal T}(T(F_{\theta^*}),T(F_{\theta}))\le \zeta,
$$
and then from (\ref{eq:Bd_T}), (\ref{eq:Hforconcfixedarea})
\beeq
\label{eq:concentration1}
 P_{\theta^*}^{(n)}(H\cap B_{\epsilon_n }) \ge  P_{\theta^*}^{(n)} [d_{{\cal F}}(F_{\theta^*}, F_{\theta}) \le \tilde \epsilon
\cap  B_{\epsilon_n }].
\eneq
Since
\beeq
\label{eq:triang1}
d_{\cal F}(F_{\theta^*},F_{\theta})\le d_{{\cal F}}(F_{\theta^*}, \hat F_{{\bf x}^*})+
d_{{\cal F}}(\hat F_{{\bf x}^*}, \hat F_{{\bf x}}) +d_{{\cal F}}(\hat F_{{\bf x}},  F_{\theta})
\eneq
if 
$$d_{{\cal F}}(F_{\theta^*}, \hat F_{{\bf x}^*})+
d_{{\cal F}}(\hat F_{{\bf x}^*}, \hat F_{{\bf x}}) +d_{{\cal F}}(\hat F_{{\bf x}},  F_{\theta}) \le \tilde \epsilon 
\mbox{ then } d_{\cal F}(F_{\theta^*},F_{\theta})\le \tilde \epsilon$$
and therefore, for the right side of (\ref{eq:concentration1})
\beeq
\label{eq:concentration2}
 P_{\theta^*}^{(n)} [d_{{\cal F}}(F_{\theta^*}, F_{\theta}) \le \tilde \epsilon
\cap  B_{\epsilon_n }) \ge  P_{\theta^*}^{(n)} [d_{{\cal F}}(F_{\theta^*}, \hat F_{{\bf x}^*})+ d_{{\cal F}}(\hat F_{{\bf x}^*}, \hat F_{{\bf x}}) +d_{{\cal F}}(\hat F_{{\bf x}},  F_{\theta}) \le \tilde \epsilon  \cap  B_{\epsilon_n }).
\eneq
From the assumptions, 
$$d_{{\cal F}}(F_{\theta^*}, \hat F_{{\bf X}^*}) \le \frac{o(k_n)}{k_n} \mbox{ and } d_{{\cal F}}(F_{\theta}, \hat F_{{\bf X}}) \le \frac{o(k_n)}{k_n}$$
with $P_{\theta^*}^{(n)}$ and $P_{\theta}^{(n)}$ probabilities converging to one, respectively, and assuming ${\bf x}^*, {\bf x}$
are in these subsets 
 the right side of (\ref{eq:concentration2})
\beeq
\label{eq:concentration3}
 P_{\theta^*}^{(n)} [d_{{\cal F}}(F_{\theta^*}, \hat F_{{\bf x}^*})+ d_{{\cal F}}(\hat F_{{\bf x}^*}, \hat F_{{\bf x}}) +d_{{\cal F}}(\hat F_{{\bf x}},  F_{\theta}) \le \tilde \epsilon  \cap  B_{\epsilon_n })
\ge  P_{\theta^*}^{(n)} [d_{{\cal F}}(\hat F_{{\bf x}^*}, \hat F_{{\bf x}}) \le \tilde \epsilon -2 \frac{o(k_n)}{k_n} \cap B_{\epsilon_n}].
\eneq
For  $\epsilon_n \downarrow 0$  as $n$  increases, eventually
\beeq
\label{eq:entildee}
\epsilon_n \le \tilde \epsilon -\frac{2o(k_n)}{k_n},
\eneq
 and the right side of (\ref{eq:concentration3})
\beeq
\label{eq:concentration4}
 P_{\theta^*}^{(n)} [d_{{\cal F}}(\hat F_{{\bf x}^*}, \hat F_{{\bf x}}) \le \tilde \epsilon -2 \frac{o(k_n)}{k_n} \cap B_{\epsilon_n}]= P_{\theta^*}^{(n)} [B_{\epsilon_n}].
\eneq
(\ref{eq:concfixedarea}) follows from (\ref{eq:concentration1}), (\ref{eq:concentration2})-(\ref{eq:concentration4}) since,
 when taking  the limit    in  (\ref{eq:ABCposteriorprob})   as $n$ increases to
infinity,  for large $n$ numerator and denominator coincide. \hspace{5ex}$\Box.$

\begin{center}
{\bf Acknowledgments}
\end{center}

Many thanks are due to Professor Rudy Beran for communicating pertinent useful  results in his 1986 paper with Professor Warry  Millar.

\pagebreak

\begin{figure}[p]
\centering
\includegraphics[width=1.15\textwidth, height=.6\textheight]{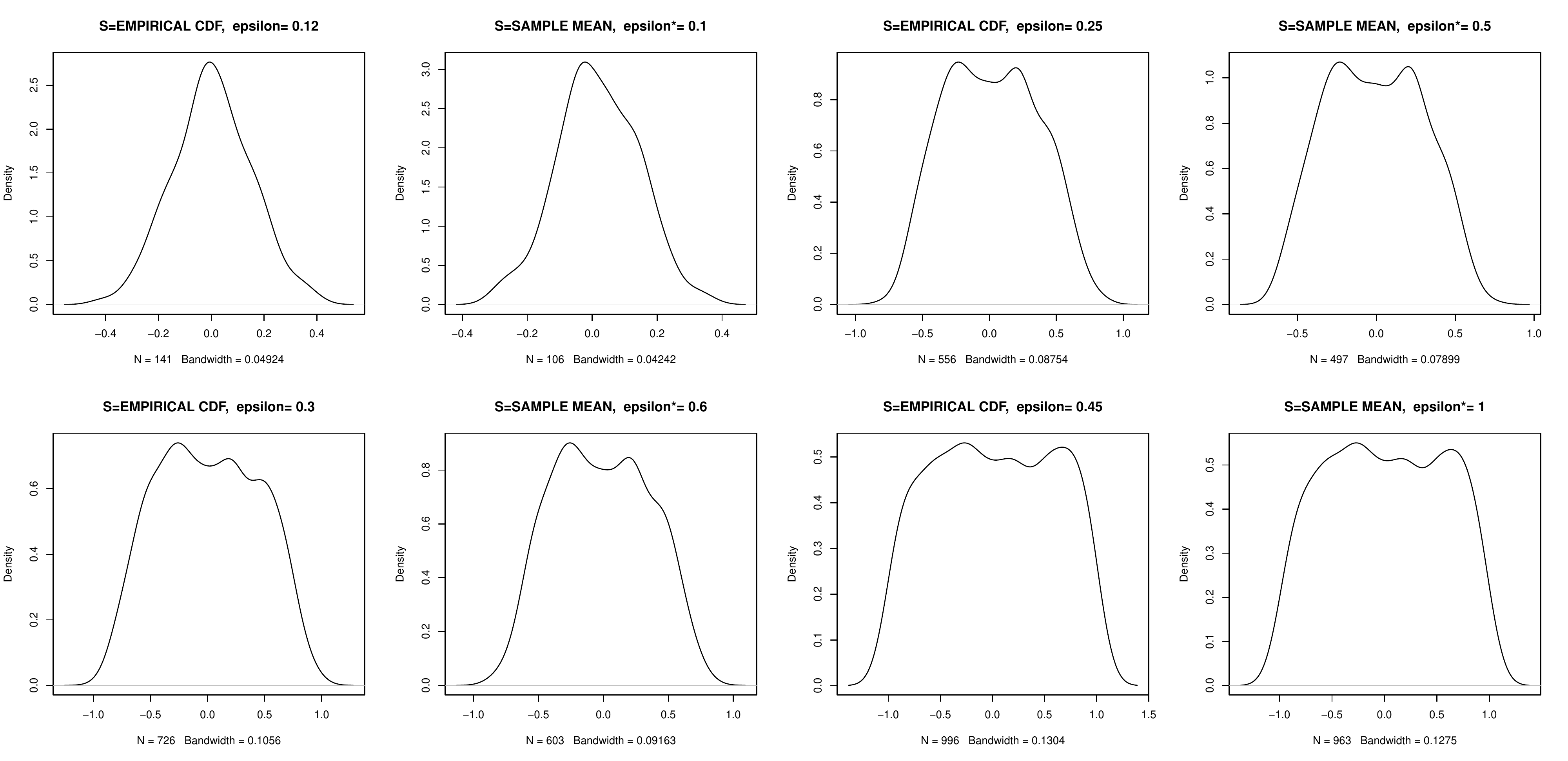}
\caption{Approximate posterior densities for various tolerance levels}
\label{fig:method}
\end{figure}
 
\pagebreak

\begin{figure}[p]
\centering
\includegraphics[width=1.15\textwidth, height=.6\textheight]{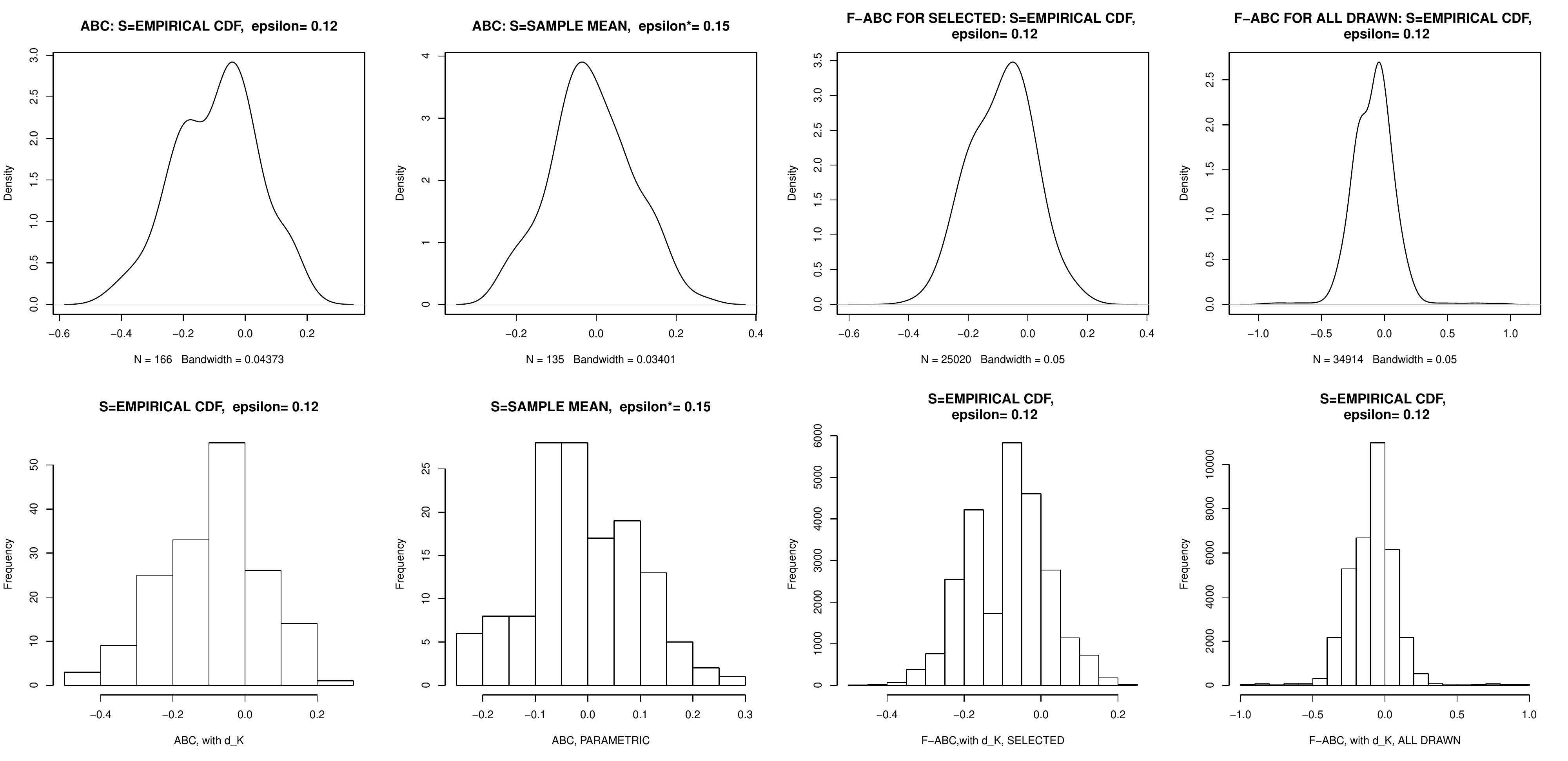}
\caption{Approximate posterior densities and histograms for ABC and F-ABC \# 1}
\label{fig:method}
 \end{figure}

\pagebreak

 \begin{figure}[p]
\centering
\includegraphics[width=1.15\textwidth, height=.6\textheight]{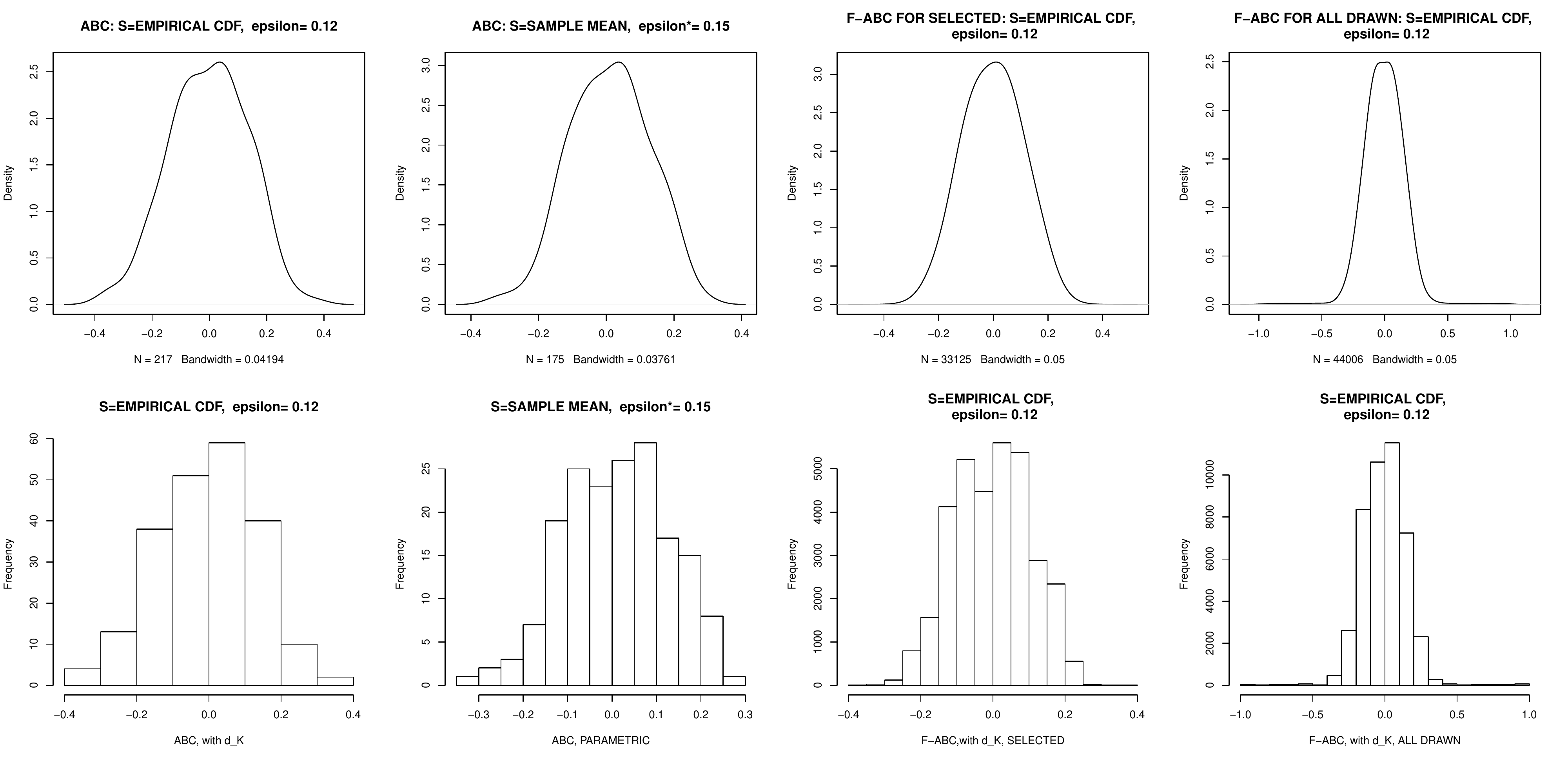}
\caption{Approximate posterior densities and histograms for ABC and F-ABC \# 2}
\label{fig:method}
 \end{figure}

\pagebreak

\begin{figure}[p]
\centering
\includegraphics[width=1.15\textwidth, height=.9\textheight]{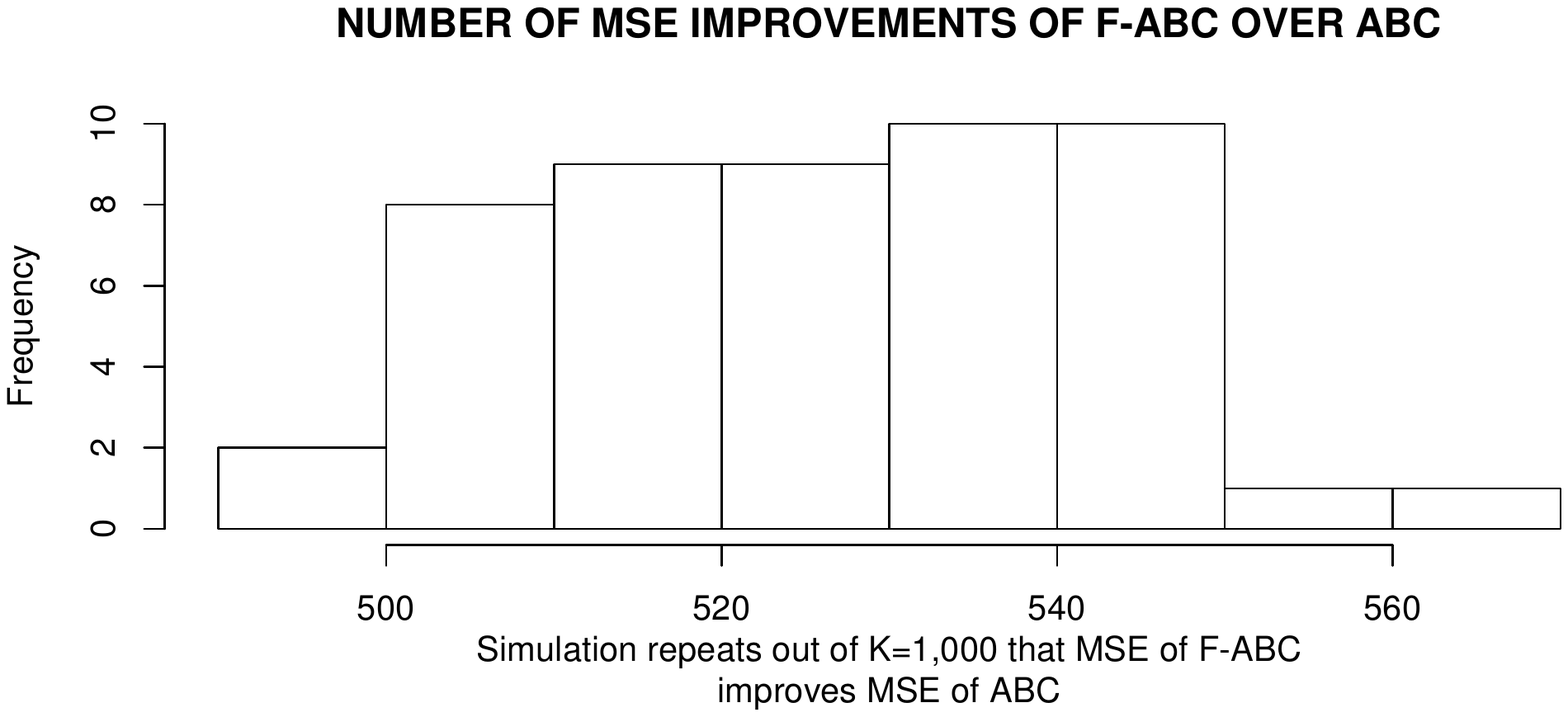}
\caption{F-ABC improves the  MSE of ABC more than 90\% of the time.}
\label{fig:method}
 \end{figure}

\pagebreak

\begin{figure}[p]
\centering
\includegraphics[width=1.15\textwidth, height=.95\textheight]{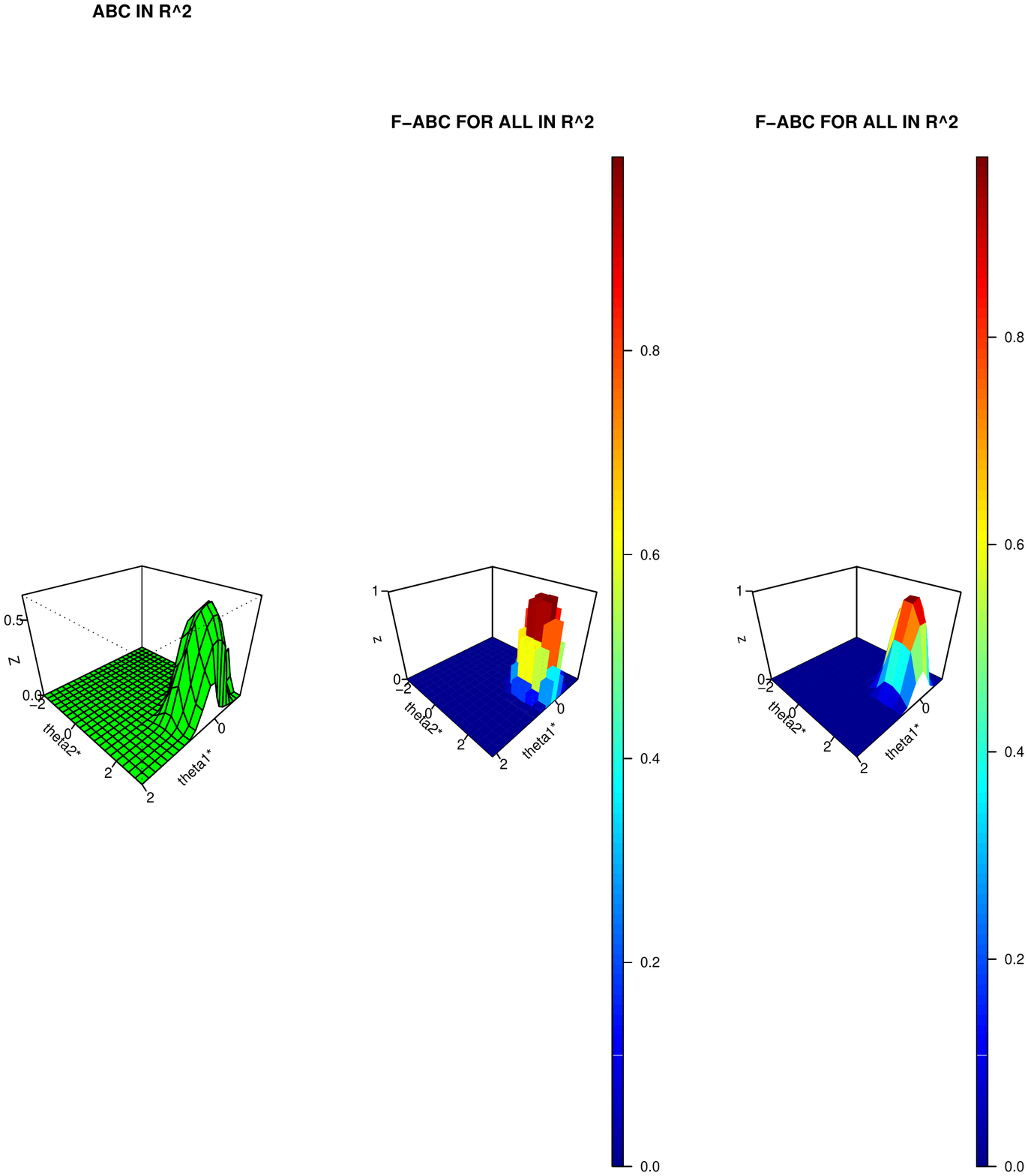}
 \caption{$d_K$ on discretized half-spaces for ABC and F-ABC for all. ABC: weights from  default kernel in $R.$
F-ABC for all: weights from  repeated samples. K=50 NS=15 M=200}
\label{fig:method}
 \end{figure}

\pagebreak

\begin{figure}[p]
\centering
\includegraphics[width=1.15\textwidth, height=.95\textheight]{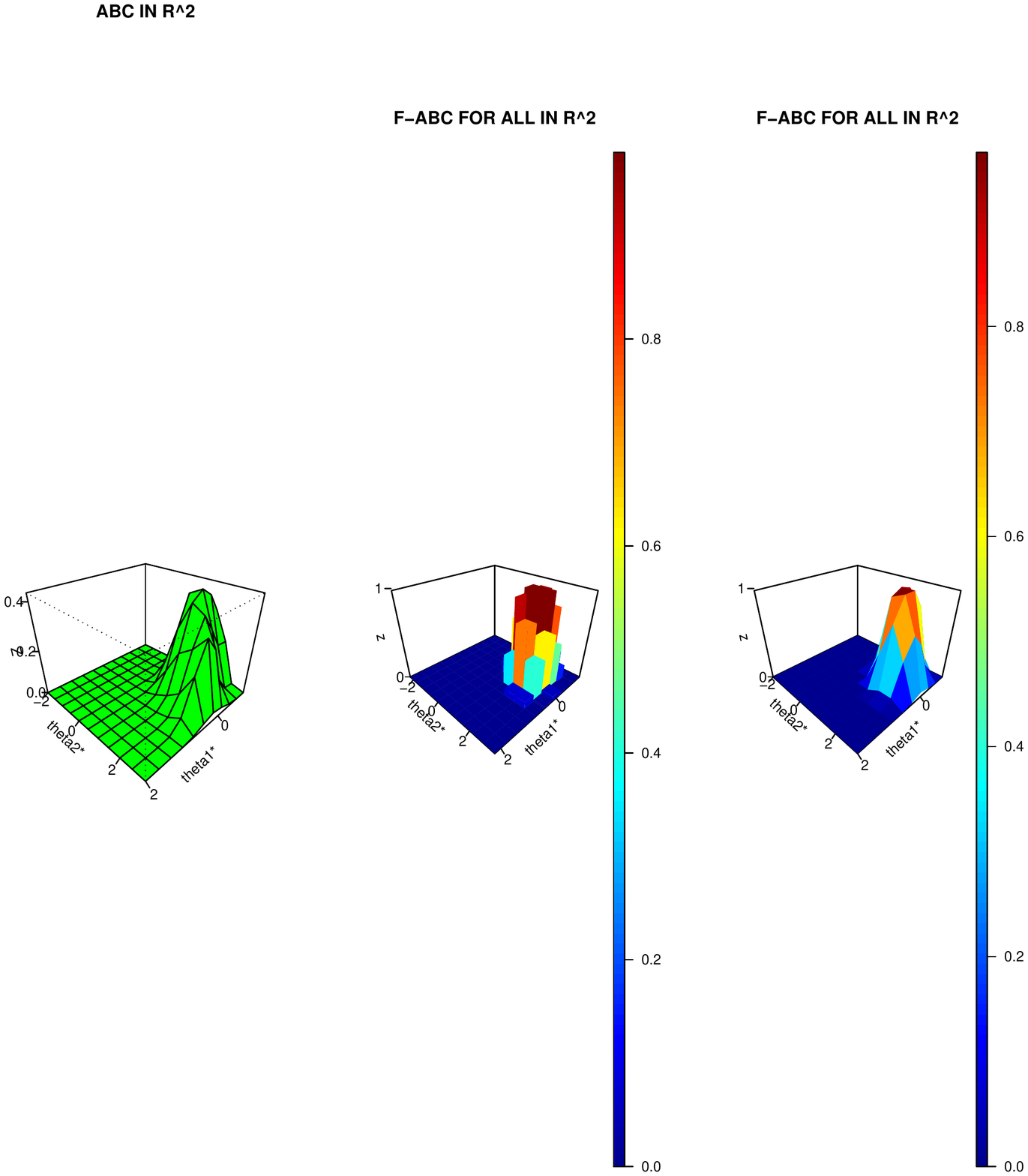}
\caption{$d_K$ on discretized half-spaces for ABC and F-ABC for all. ABC: weights from  default kernel in $R.$
 F-ABC for all: weights from  repeated samples.  K=10 NS=10 M=50}
\label{fig:method}
 \end{figure}


 


\begin{thebibliography}{xxx99}

\bibitem{xyz} Beran, R. and Millar, P. W. (1986) Confidence Sets for a Multivariate Distribution. {\em Ann. Statist.}
{\bf 14}, 431-443.

\bibitem{xyz} 
Bornn, L., Pillai, N. S., Smith, A. and Woodard, D. (2017) The Use of a Single Pseudo-Sample
in Approximate Bayesian Computation. {\em Stat. and Comput.} {\bf 27}, 583-590.

\bibitem{xyz} Bernton, E. , Jacob, P. E., Gerbery, M. and  Robert, C. P.
(2019)  Approximate Bayesian computation with the Wasserstein
distance. arXiv:1905.03747v1
 
\bibitem{xyz} Biau, G., C\'erou, F. and  Guyader, A. (2015) New insights into approximate Bayesian computation. {\em Annales de l' IHP (Probab. Stat.)}  {\bf 51}, 376-403 

\bibitem{xyz} Breiman, L. (1992) {\em Probability} Classics  in Applied Mathematics,  SIAM.

\bibitem{xyz} Cram\'er, H. and Wold, H. (1936) Some theorems on distribution functions.{\em J. London Math. Soc.} {\bf 11}, 290-294.


\bibitem{xyz} Devroye, L. P. (1977) A Uniform Bound for the Deviation of Empirical Distribution Functions. {\em J. Multiv. Anal.}
{\bf 7}, 594-597.

\bibitem{xyz}  Dudley, R. M. (1984) A course on empirical processes. \'Ecole d' \'Et\'e de Probabilit\'es de St. Flour. {\em Lecture Notes in Math.} {\bf 1097}, 2-142, Springer Verlag, New York.



\bibitem{xyz}  Dudley, R. M. (1978)  Central limit theorem for empirical measures. {\em Ann. Prob.} {\bf 6}, 899-929.

\bibitem{xyz}  Dvoretzky, A.,  Kiefer, J. and  and Wolfowitz, J. (1956) Asymptotic minimax character of the sample distribution function and of the classical multinomial estimator. {\em Ann. Math. Stat.} {\bf 27}, 642-669
 
\bibitem{xyz} Fearnhead, P. (2018) Asymptotics of ABC.  {\em Handbook of Approximate Bayesian Computation}, Editor:Routledge Handbooks Online.

\bibitem{xyz}  Fearnhead, P.  and  Prangle, D. (2012) Constructing summary statistics for approximate Bayesian computation: semi-automatic approximate Bayesian computation. {\em J. R. Statist. Soc. B} {\bf 74}, 419-474.


\bibitem{xyz} Frazier, D. T., Martin, G. M., Robert, C. P. and Rousseau, J. (2018) Asymptotic properties  of approximate Bayesian Computation. {\em Biometrika} {\bf 105}, 593-607.

\bibitem{xyz} Frazier, D. T., Martin, G. M. and Robert, C. P. (2015) On Consistency of Approximate Bayesian Computation arXiv:1508.05178v1

\bibitem{xyz}   Kiefer, J. (1961)  On Large Deviations of the Empiric D. F. of Vector Chance Variables and a Law of the Iterated logarithm. {\em Pacific J. of Mathematics} {\bf 11}, 649-660

\bibitem{xyz}   Kiefer, J. and Wolfowitz, J. (1958) On the deviations of the empiric distribution function of vector chance variables.
{\em Trans. Amer. Math. Soc. } {\bf 87}, 173-186

\bibitem{xyz}  Lauritzen, S.  (2007)  Exchangeability and de Finetti's Theorem. Lecture Notes, University of Oxford,\\
http://www.stats.ox.ac.uk/$\sim$steffen/teaching/grad/definetti.pdf

\bibitem{xyz}  Lintusaari, J., Gutmann, M. U., Dutta, R., Kaski, S. and Corander, J. (2017) Fundamentals and Recent Developments in Approximate Bayesian Computation. {\em Syst. Biol.} {\bf 66}, e66-e82.

\bibitem{xyz} Massart, P. (1990) The tight constant in the Dvoretzky-Kiefer-Wolfowitz inequality. {\em Ann. Prob.} {\bf 18}, 1269-1283

\bibitem{xyz}   Miller, J. W. and Dunson, D. B. (2019)   Robust Bayesian inference via coarsening. {\em J. Am. Stat. Assoc. }
{\bf 114}, 1113-1125

\bibitem{xyz} Nott, D. J., Drovandi, C. C., Mengersen, K. and  Evans, M. (2018) Approximation of Bayesian Predictive p-Values with Regression ABC. {\em Bayesian Analysis} {\bf 13}, 59-83.

\bibitem{xyz}  Pritchard, J. K., Seilstad, M. T.,  Perez-Lezaun, A and Feldman, M. W. (1999)
 Population Growth
of Human Y Chromosomes: A Study of Y Chromosome Microsatellites. {\em Molecular Biology and Evolution}, {\bf 16},
1791-1798.
 

\bibitem{xyz} Robert C.P. (2016) Approximate Bayesian Computation: A Survey on Recent Results. In: Cools R., Nuyens D. (eds) Monte Carlo and Quasi-Monte Carlo Methods. Springer Proceedings in Mathematics \& Statistics, vol 163. Springer, Cham

\bibitem{xyz} Rubin, D. B. (2019) Conditional Calibration and the Sage Statistician. {\em Survey Methodology} {\bf 45}, 187-198.
 
\bibitem{xyz} Rubin, D. B. (1984) Bayesianly Justifiable and Relevant Frequency Calculations for the Applied Statistician. {\em Ann.
Statist.} {\bf 12}, pp. 213-244.

\bibitem{xyz}  
 Tanaka, M.M.,   Francis, A. R., Luciani, F. and  Sisson, S. A. (2006) 
Using Approximate Bayesian Computation to Estimate Tuberculosis
Transmission Parameters From Genotype Data. {\em Genetics}, {\bf 173}, 1511–1520.

\bibitem{xyz} Tavar\'{e}, S.  (2019).  An introduction to Approximate Bayesian Computation. {\em Summer Program, Herbert and 
Florence Irving Institute for Cancer Dynamics.} https://cancerdynamics.columbia.edu/content/summer-program


\bibitem{xyz} Tavar\'{e}, S., Balding, D. J., Griffiths, R. C.  and Donnelly, P. (1997). Inferring Coalescence Times from
DNA Sequence Data, Genetics, 145, 505-518.

\bibitem{xyz} Vihola, M. and Franks, J. (2020) On the use of approximate Bayesian  computation Markov chain Monte Carlo with inflated tolerance and post correction. {\em Biometrika} https://doi.org/10.1093/biomet/asz078 
 

\bibitem{xyz} Yatracos, Y. G.   (2020) Matching Estimation for Data Generating Experiments.
Preprint





 
\end{thebibliography}
\end{document}